\documentclass{cernrep}
\usepackage{texnames}
\usepackage[T1]{fontenc}
\usepackage{subfig}

\graphicspath{{/Figures/}{/figures/}}

\begin{document}

\title{Double-Frame Current Control with a Multivariable PI Controller and Power Compensation for Weak Unbalanced Networks}

\author{D. Siemaszko$^*$ and A. Rufer$^\dag$}
\institute{$^*$ PESC-CH (Power Electronics \& Systems Consultancy Switzerland), Geneva, Switzerland \\
$^\dag$ EPFL (Ecole Polytechnique F\'ed\'erale de Lausanne), Lausanne, Switzerland}

\maketitle

\begin{abstract}
The handling of weak networks with asymmetric loads and disturbances implies the accurate handling of the second-harmonic component that appears in an unbalanced network. This paper proposes a classic vector control approach using a PI-based controller with superior decoupling capabilities for operation in weak networks with unbalanced phase voltages. A synchronization method for weak unbalanced networks is detailed, with dedicated dimensioning rules. The use of a double-frame controller allows a current symmetry or controlled imbalance to be forced for compensation of power oscillations by controlling the negative current sequence. This paper also serves as a useful reminder of the proper way to cancel the inherent coupling effect due to the transformation to the synchronous rotating reference frame, and of basic considerations of the relationship between switching frequency and control bandwidth.
\end{abstract}


\section{Introduction}
\label{sec:intro}

In the field of grid-connected power converters that interface a DC link to an AC three-phase network, the control of the current taken or injected presents challenges that are not taken into account when infinite networks are considered. In fact, networks are becoming weaker and weaker owing to the multiplication of decentralized power sources and islanded systems. As a consequence, there is a strong need for a synchronization system that can handle the disturbances inherent in weak networks, and the harmonics produced in the controlled power converter itself.

Moreover, the need for operation in an unbalanced network implies a need to control the second-harmonic component that appears more generally when an asynchronous network is operated in a synchronous reference frame. Also, being generally limited in bandwidth because of the very low switching frequencies used on the network side, the current controller is unable by itself to compensate for second-harmonic perturbation.

A common solution, not focused on any particular application, must be found for any grid converter connected to a weak network \cite{pll.siemaszko}. This paper offers a simple method that is applicable over a wide range of fields, including industrial drives, wind generators, ship applications such as hybrid propulsion and onboard network supplies, electric vehicles (with support for smart grids), aircraft AC and DC onboard networks, and any islanded system, moving or not.

\section{Converter and grid considerations}
\label{sec:grid}

\begin{figure}[!ht]
		\centering
    \includegraphics[scale=0.5]{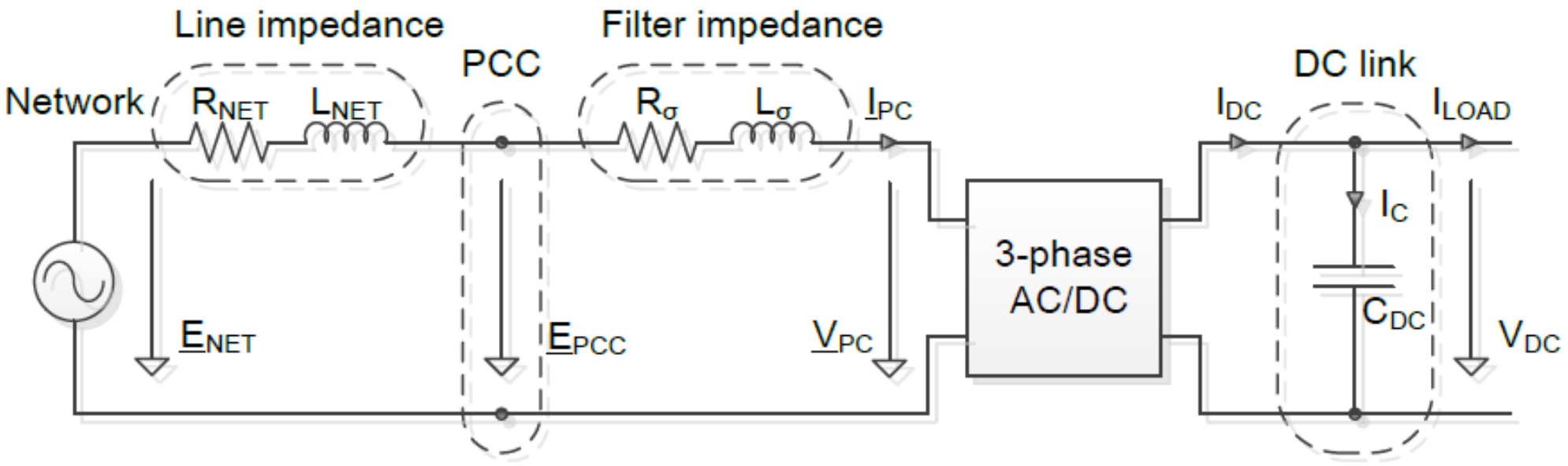}
    \caption{Grid-connected power converter on system level}
    \label{g.converter_model}
\end{figure}

The typical voltage source inverter connects a DC link to the grid through a transformer with a filter on its secondary side. As illustrated in Fig. \ref{g.converter_model}, the grid transformer may be modelled by its leakage impedance added to the filter inductance. The main purpose of the converter is to control the current, either to maintain the DC link voltage $V_\mathrm{DC}$ or to maintain the voltage on the network side. The interfaced signals and system parameters are described in Table \ref{t.system_signals}. The network as seen from the power converter side may be modelled as a voltage source $V_\mathrm{NET}$ and its grid impedance $Z_\mathrm{NET}$ as seen from the converter side, reflecting its strength as in Ref. \cite{rl.cobreces}.

The converter is operated using a vector control strategy. The grid voltages $E_\mathrm{PCC}$ measured at the point of common coupling and the converter currents $I_\mathrm{PC}$ can be described in the synchronous reference frame computed by the well-known Park transformation; here, proper decoupling with respect to network perturbations is a key requirement for a stable controller. The synchronization with the grid, which is done with a dedicated phase-locked loop (PLL) that provides the reference for the vector controller, is detailed in the following section.

\begin{table}[h!]
		\centering
		\caption{List of signals and parameters in the system}
		\label{t.system_signals}
		\begin{tabular}{rc|l}
\hline\hline
		{\bf Signal} & {\bf Units} & {\bf Description} \\
		\hline
		$\underline{E}_\mathrm{NET}$ &  V  			& Three-phase network voltage\\
		$\underline{E}_\mathrm{PCC}$ &  V  			& Three-phase voltage measured at PCC\\
		$\omega_\mathrm{N}$ 					&  Hz  			& Angular frequency of the synchronous rotating frame\\
		$\theta_\mathrm{N}$ 					&  rad  		& Position angle of the synchronous rotating frame\\
		$\underline{V}_\mathrm{PC}$ 	&  V  			& Power converter voltage on grid side\\
		$\underline{I}_\mathrm{PC}$ 	&  A  			& Power converter current on grid side\\
		$V_\mathrm{DC}$ 							&  V  			& DC link voltage\\
		$I_\mathrm{DC}$ 							&  A  			& Feeding current from grid converter\\
		$I_\mathrm{C}$ 							&  A  			& DC link capacitor current\\
		$I_\mathrm{LOAD}$ 						&  A  			& Load current\\
		&&\\
		
\hline
{\bf Parameter} & {\bf Units} & {\bf Description} \\
		\hline
		$C_\mathrm{DC}$ 							&  F 				& DC link capacitance\\
		$L_\sigma$ 						&  H 				& Equivalent line inductance on the converter side\\
		$R_\sigma$ 						& $ \Omega $	& Equivalent line resistance on the converter side\\
		$Z_\sigma$ 						& $ \Omega $	& Equivalent total line impedance on the converter side\\
		$L_\mathrm{NET}$ 						&  H 				& Equivalent line inductance of the network\\
		$R_\mathrm{NET}$ 						& $ \Omega $	& Equivalent line resistance of the network\\	
		$Z_\mathrm{NET}$ 						& $ \Omega $	& Equivalent total line impedance of the network\\
		$f_\mathrm{NET}$							&  Hz 			& Frequency of the network\\
		$\phi_\mathrm{NET}$					&  rad 			& Angle of equivalent network impedance phasor\\
		$S_\mathrm{SCC}$ 						&  VA 			& Short-circuit power of the network\\\hline\hline
		\end{tabular}
\end{table}

\subsection{Strength of the grid}
\label{sub:strength}

\begin{figure}[!h]
		\centering
    \includegraphics[scale=0.5]{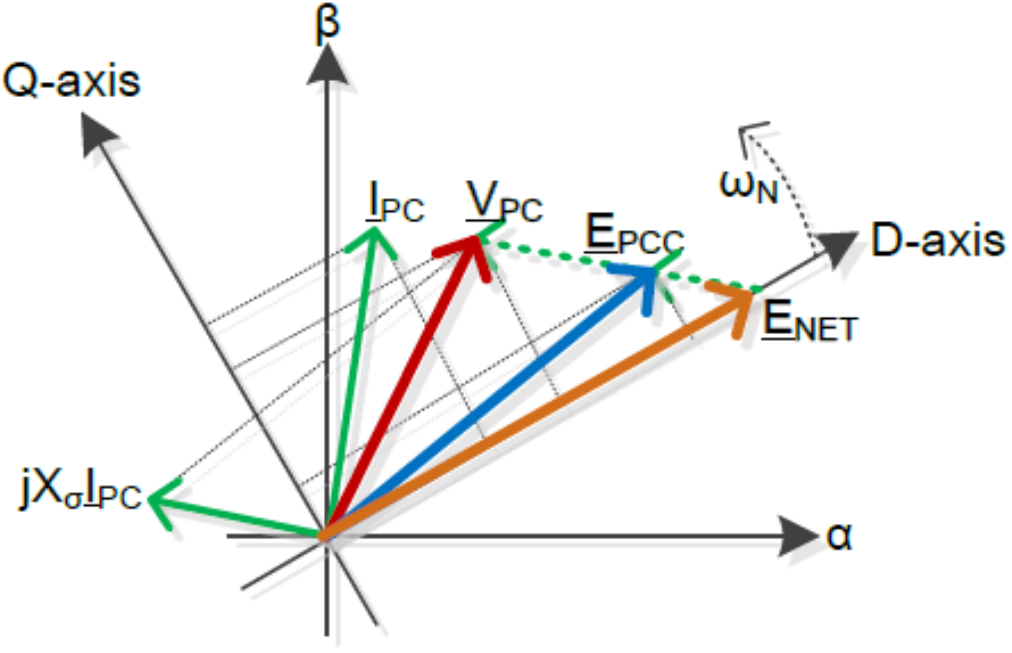}
    \caption{Principle of network synchronization and vector control}
    \label{g.Voltage_vectors}
\end{figure}

\begin{figure}[!h]
\centering
	\subfloat[][]{\includegraphics[width=0.25\linewidth]{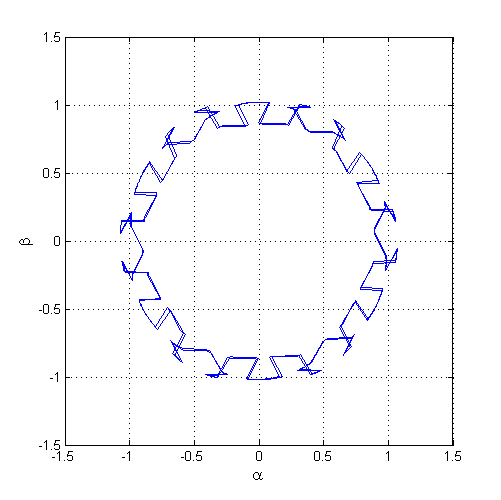}\label{g.PCC}}
	\subfloat[][]{\includegraphics[width=0.25\linewidth]{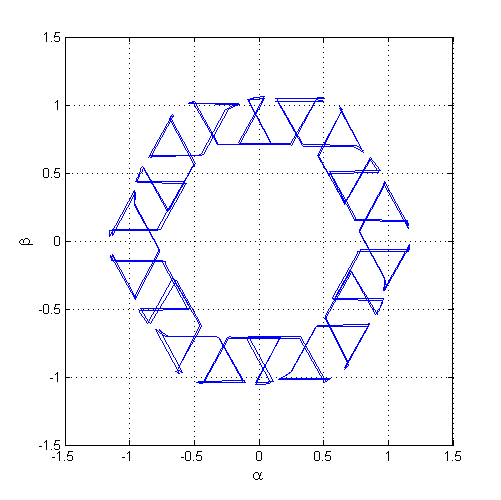}\label{g.weak_PCC}}
	\subfloat[][]{\includegraphics[width=0.25\linewidth]{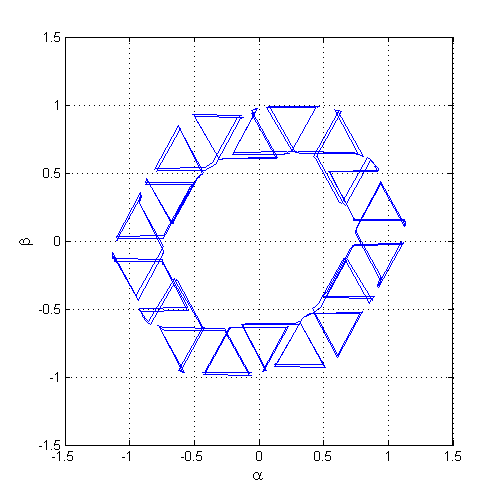}\label{g.weak_unbalance}}
	\subfloat[][]{\includegraphics[width=0.25\linewidth]{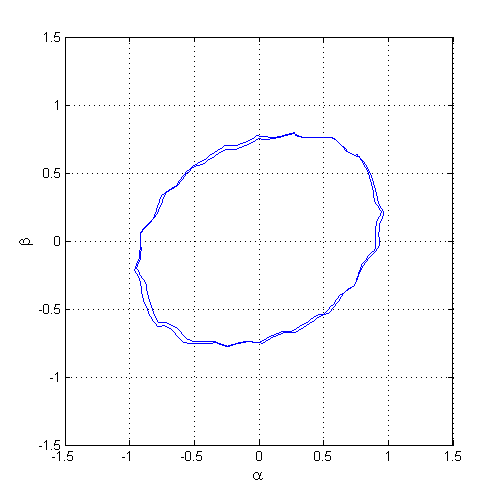}\label{g.weak_filter}}
	\caption{$\alpha\beta$ representation of voltages at PCC (a) for a strong network (20$\times$) under balanced conditions, and for a weak network (5$\times$) under (b) balanced and (c) unbalanced conditions; (d) its recomposition from decoupled filtered values.}
	\label{g.weak}
\end{figure}

The strength of a network is defined by its short-circuit power $S_\mathrm{SCC}$ as it appears at the point of common coupling (PCC). It may also be expressed as a ratio relative to the nominal power of the grid converter. Strong networks typically have a nominal power higher than 20 times that of the grid converter, and very weak networks, such as networks in ships and remote microgrids, have less than 8 times that of the grid converter. This value defines the influence of a power converter on a given network. In the case of a converter connected to a weak network, $Z_\mathrm{NET}$ is not negligible compared with $Z_\sigma$, and then the voltage  $E_\mathrm{PCC}$ as measured at the point of common coupling is different from the idealized network voltage $V_\mathrm{NET}$, as illustrated in Fig. \ref{g.Voltage_vectors}. In such a case, harmonics injected by the power converter appear in the measured voltages, which significantly disturb the controller as illustrated in
\Fref{g.weak}\subref{g.weak_PCC}.
Under conditions of phase voltage imbalance, the approximate circle in the Cartesian plane becomes an ellipse as in
\Fref{g.weak}\subref{g.weak_unbalance}, in which case filtering as in
\Fref{g.weak}\subref{g.weak_filter} can only be performed by applying the decoupling principles described in the following section. To assess the equivalent impedance of the network as seen from the PCC, one can use the relations given in Eq.~(\ref{equ.znet}):
\begin{equation}
\begin{cases}
Z_\mathrm{NET} = \cfrac{E_\mathrm{NET}^{2}}{S_\mathrm{SCC}},\\
R_\mathrm{NET}	= Z_\mathrm{NET}\cos{\phi_\mathrm{NET}},\\
L_\mathrm{NET}	= \cfrac{Z_\mathrm{NET}\sin{\phi_\mathrm{NET}}}{2\pi f_\mathrm{NET}}.
\end{cases}\label{equ.znet}
\end{equation}

\section{Grid synchronization}
\label{sec:sync}

The most common way of synchronizing a controller with an alternating signal is to use a PLL. The quadrature part of the network voltage, transformed into the synchronous reference frame, is used as the rotating-frame phase error input to a PI controller. However, the use of simple filters is not adequate for synchronization with weak networks with harmonics and asymmetric disturbances. Several methods for improving the robustness of the synchronization of the system with the network have been presented \cite{pll.BenhabibEPE,pll.RoblesEPE,pll.suulEPE,pll.rodriguezEPE}. In the work presented in this paper, the method implemented for decoupling the second harmonic inherent in a voltage imbalance was inspired by Ref. \cite{pll.rodriguez}, where decoupling of the positive and negative sequences of the voltage allows all perturbations in each reference frame to be properly filtered, and then the sequences are recombined to obtain a clean representation of the network with selected harmonics.

\subsection{Phase-locked loop}
\label{sub:pll}

\begin{figure}[!ht]
		\centering
    \includegraphics[scale=0.5]{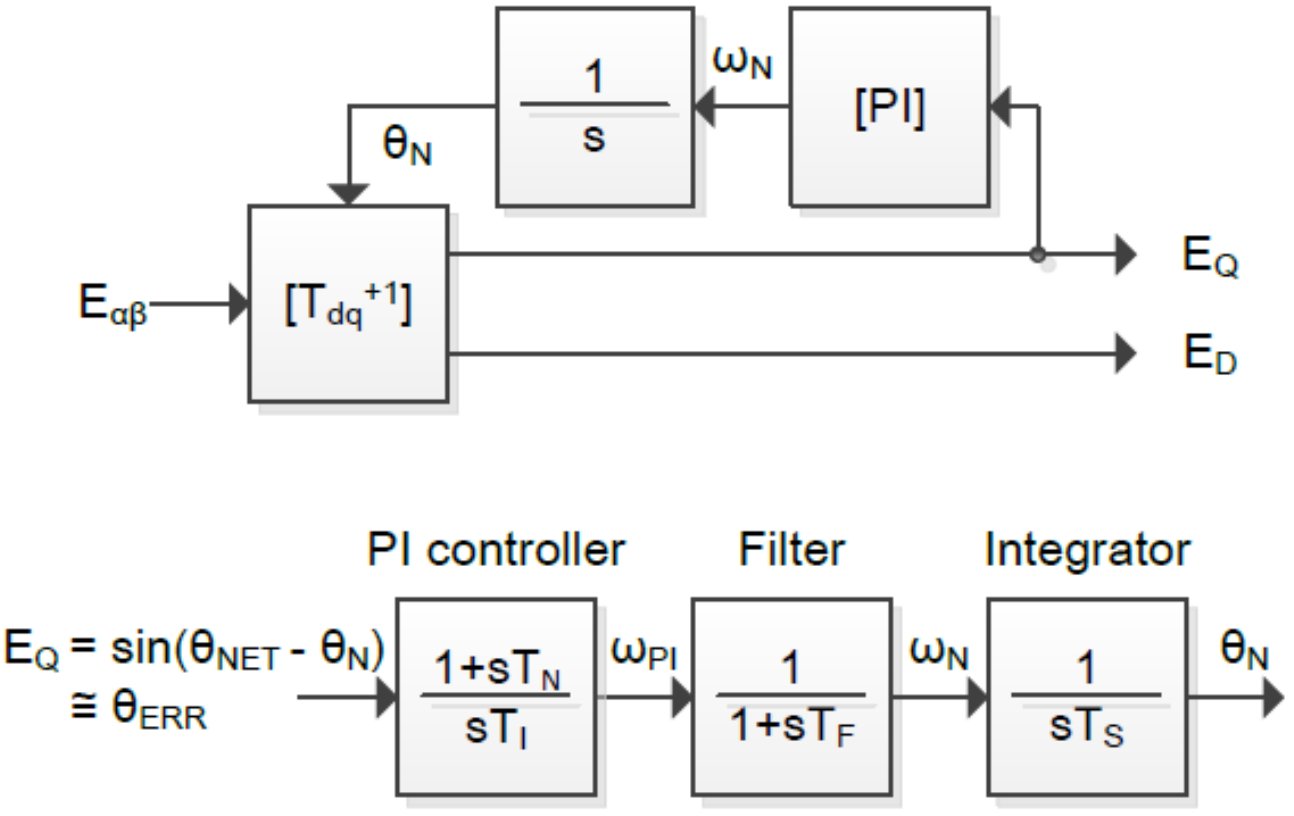}
    \caption{Equivalent control structure of the PLL}
    \label{g.pll_ctrl}
\end{figure}

The control principle of the synchronous-reference-frame PLL was based on the minimization of the phase error between the rotating frame and the network, as illustrated in Fig. \ref{g.pll_ctrl}. The quadrature component of the voltage was used directly as the phase error, since one can take the sine of a small quantity as almost equal to that quantity. Using a simple PI controller, the resulting value was filtered to obtain the state variable $\omega_\mathrm{N}$, which was integrated to obtain the phase function $\theta_\mathrm{N}$.

The parameters of the PI controller were set using a magnitude optimum criterion given by the transfer functions of the filter, $G_\mathrm{F}(s)$ (Eq. (\ref{equ.PLLgf})), the plant, $G_\mathrm{S} (s)$ (Eq. (\ref{equ.PLLgs})) (which was actually the integrator), and the controller itself, $G_\mathrm{R} (s)$, as described in {Eq. (\ref{equ.PLLgr}). The time constant $T_\mathrm{N}$ of the controller was set in order to compensate the dominant time constant of the system $T_\mathrm{S}$ (Eq. (\ref{equ.PLLtn})), and the integral time constant $T_\mathrm{I}$ was set as a function of the filter time constant $T_\mathrm{F}$ (Eq. (\ref{equ.PLLti})):
\begin{eqnarray}
G_\mathrm{F}(s) & = & \cfrac{1}{1 + sT_\mathrm{F}}, \label{equ.PLLgf}\\
G_\mathrm{S} (s) & = &\cfrac{1}{sT_\mathrm{S}}, \label{equ.PLLgs}\\
G_\mathrm{R} (s) & = & \cfrac{1 + sT_\mathrm{N}}{sT_\mathrm{I}}, \label{equ.PLLgr}\\
T_\mathrm{N}	& =	& T_\mathrm{S} = \cfrac{1}{\omega_{NET}}, \label{equ.PLLtn}\\
T_\mathrm{I}	& =	& 2K_\mathrm{F} K_\mathrm{S} T_\mathrm{F} = 2T_\mathrm{F}. \label{equ.PLLti}
\end{eqnarray}

As a result, the open-loop transfer function $G_0(s) = G_\mathrm{R}(s)G_\mathrm{F}(s)G_\mathrm{S}(s)$ crossed the zero axis with a slope of $-$20~dB/dec and eliminated all higher frequencies, as illustrated in Fig. \ref{g.pll_bandwidth}. By computing the optimal parameters using the magnitude optimum criterion, we obtained a phase margin of $63^\circ$ that ensured stability of the system. The time constant $T_\mathrm{F}$ of the PLL filter, which defines the overall dynamics of the system, was chosen as a function of the utility grid frequency and the harmonics that we wanted to eliminate.

\begin{figure}[!ht]
		\centering
    \includegraphics[scale=0.5]{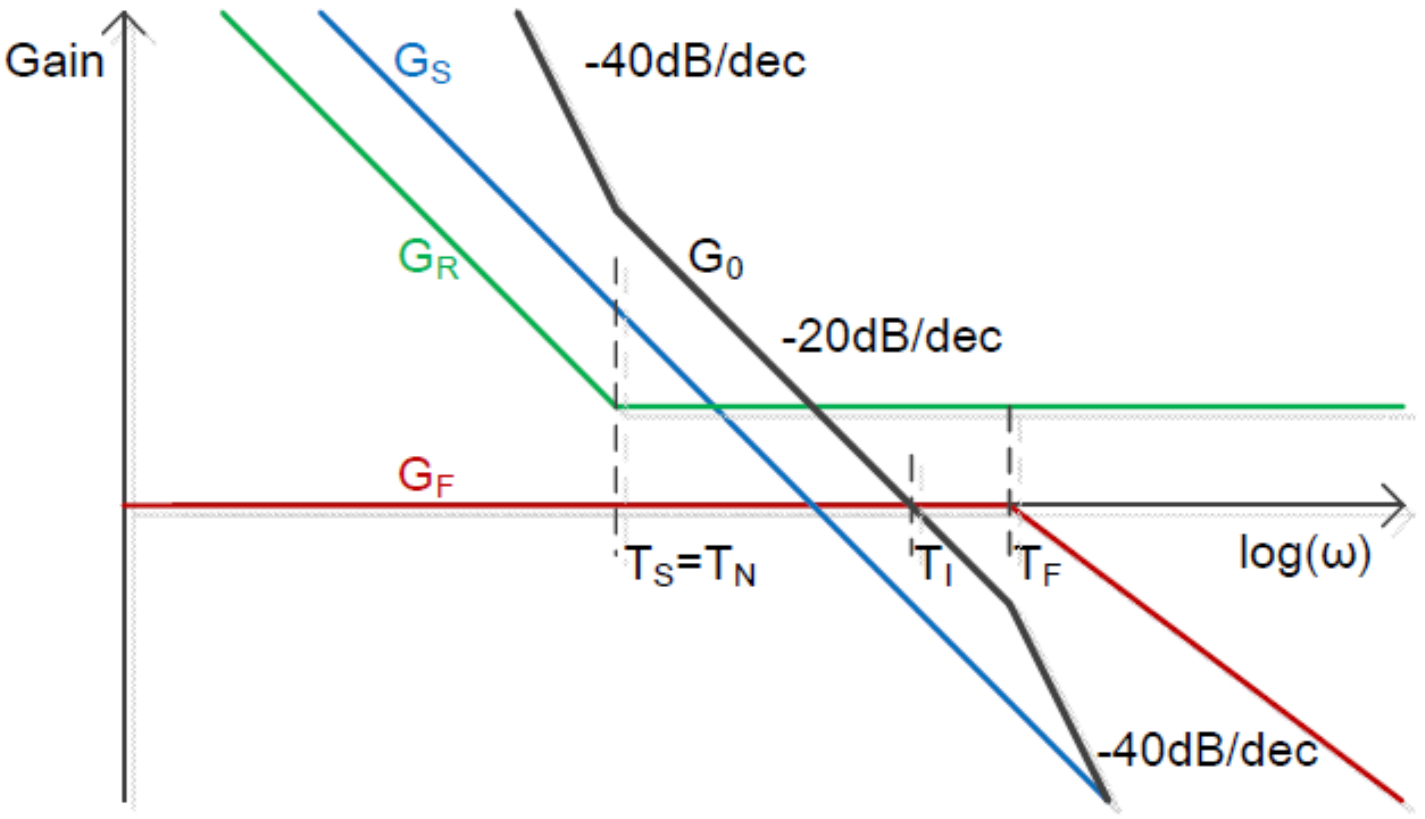}
    \caption{Bandwidth of the control structure of the PLL}
    \label{g.pll_bandwidth}
\end{figure}

\subsection{Synchronization with a weak network}

The PLL described above was tested in a hardware-in-the-loop (HIL) simulation model recreating a real industrial control platform, with its own individual sampling systems connected to a model of the power components as described in Fig. \ref{g.converter_model}. The PI controller was tested with a phase step in the phase voltages, which corresponds to a situation where a reactive load is connected to a network with which a converter must be synchronized. This phase step could be performed in the rotating frame itself to test a PLL in a real system.

As illustrated in
\Fref{g.filteringvoltage}\subref{g.Sync_weak_trans},
the PLL became synchronized with the network phase after one and a half periods, corresponding to the time constant of the PLL filter. The harmonics due to converter switching were entirely rejected as long as the phase voltages were balanced. When a phase voltage imbalance occurred, as in \Fref{g.filteringvoltage}\subref{g.Network_unbal}, the inherent second harmonic appeared in the angular frequency $\omega_\mathrm{N}$ of the reference frame, and also appeared in the phase $\theta_\mathrm{N}$.

\begin{figure}[!ht]
\centering
	\subfloat[][]{\includegraphics[width=0.35\linewidth]{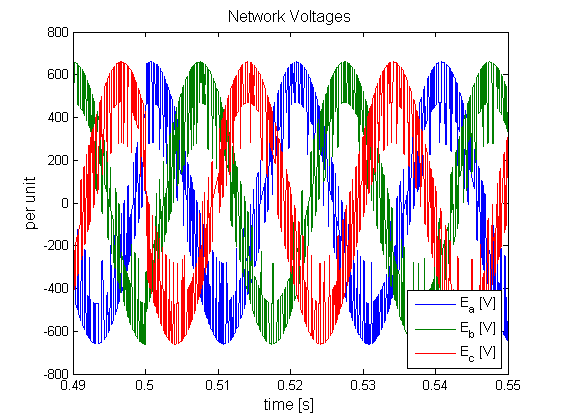}\label{g.Network_weak_trans}}
	\subfloat[][]{\includegraphics[width=0.35\linewidth]{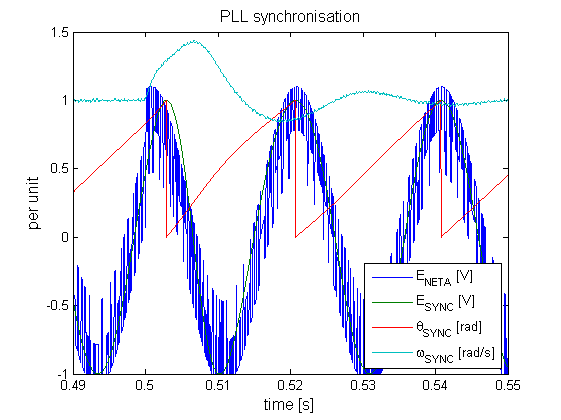}\label{g.Sync_weak_trans}}\\
	\subfloat[][]{\includegraphics[width=0.35\linewidth]{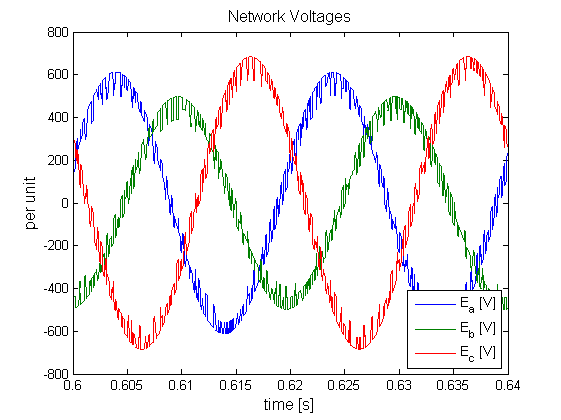}\label{g.Network_unbal}}
	\subfloat[][]{\includegraphics[width=0.35\linewidth]{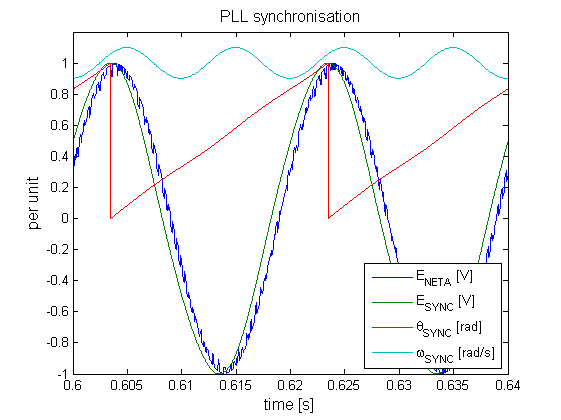}\label{g.Sync_bad}}
	\caption{(a) Phase step in the phase voltages of a weak network, (b) response of PLL to phase step, (c) asymmetric phase voltages in a strong network, and (d) effect of second harmonic in the PLL signals due to asymmetry.}
	\label{g.filteringvoltage}
\end{figure}

\subsection{Synchronization with an asymmetric network}

To operate with signals containing a strong second-harmonic content, one needs to apply a decoupling strategy in two reference frames, namely the positive- and negative-sequence synchronous rotating frames. The decoupling of the two sequences, as introduced in Ref. \cite{pll.rodriguez}, uses the core principle that the vector projection onto a reference rotating frame results in a DC component in the frequency content of a signal corresponding to that reference rotating frame. A signal containing several harmonics can be described in the Cartesian plane by the addition of several vectors of a fixed length rotating at a fixed frequency.

According to the decoupling principle illustrated in Fig. \ref{g.pll}, one can separate the reference signal into a fundamental component and its second harmonic, described individually in two different rotating reference frames. Once the two rotating vectors have been found, one can decouple them from each other in order to consider the second-harmonic perturbation separately from the fundamental harmonic.
 The benefit for a PLL that is synchronized with the rotating reference frame lies in the possibility of using a reference signal from the positive sequence decoupled from the second-harmonic perturbation. Its implementation is not as difficult as is stated in Ref. \cite{pll.carugati} and does not require the use of notch filters, which are not reliable. Moreover, the function of the decoupling is not restricted to the PLL controller; it also provides four decoupled voltage components that can be used in current control as voltage feedforward signals, and can possibly be used to compute current references for the compensation of power oscillations occurring during phase voltage asymmetries.

\begin{figure}[!ht]
		\centering
    \includegraphics[scale=0.5]{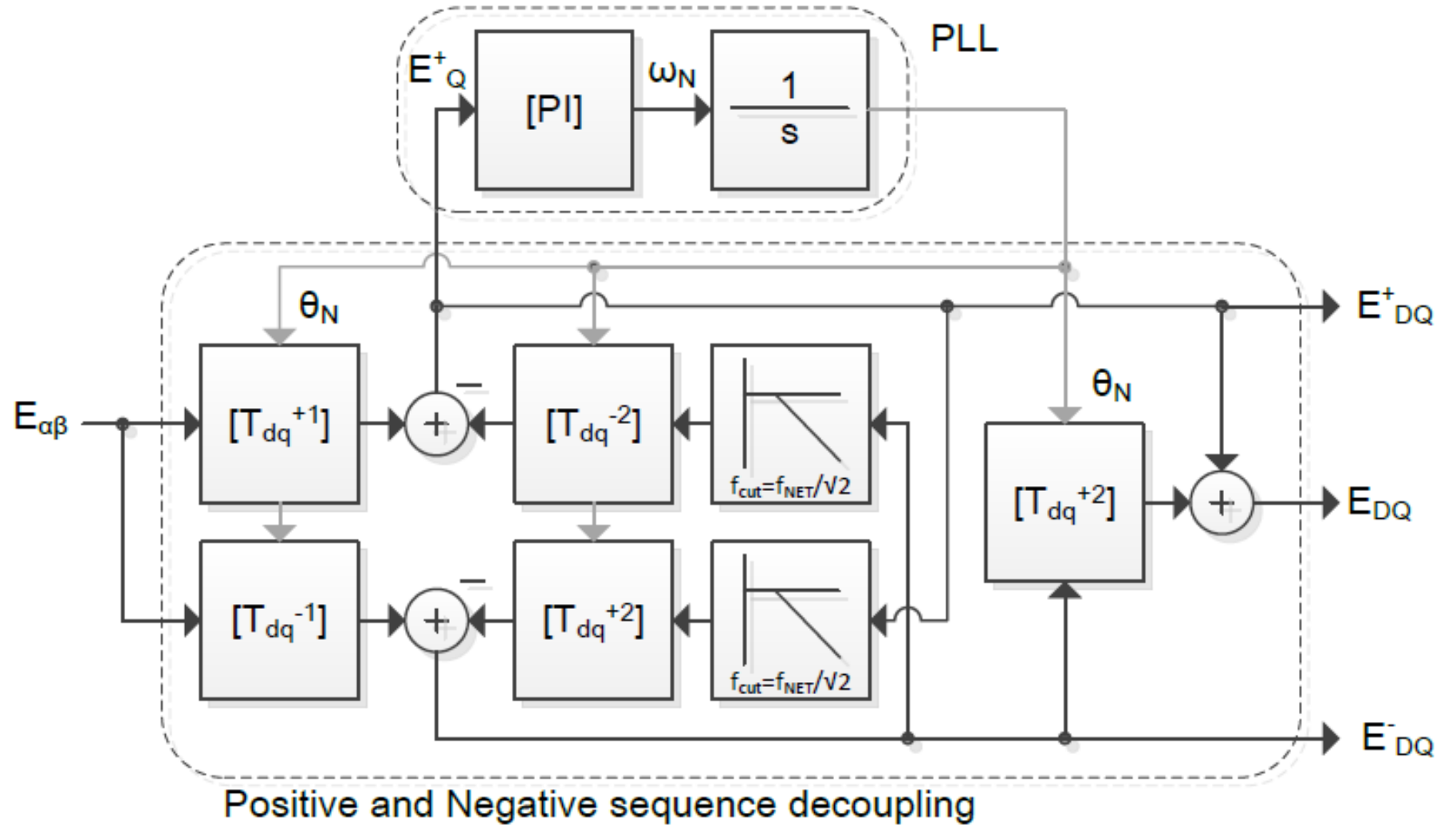}
    \caption{Double-decoupled synchronous-reference-frame PLL}
    \label{g.pll}
\end{figure}

A mathematical proof supporting this decoupling scheme can be found in Ref. \cite{pll.rodriguez}, but the overall operation principle is illustrated in Fig. \ref{g.Decoupling_illustrated}. The phase voltages are first transformed to both positive- and negative-sequence rotating reference frames, giving the four blue curves $E^+_\mathrm{DCP}$, $E^+_\mathrm{QCP}$, $E^-_\mathrm{DCP}$, and $E^-_\mathrm{Q}$. From each of them is subtracted the projection of the decoupled value from the opposite reference frame, $E^+_\mathrm{DDECNEG}$, $E^+_\mathrm{QDECNEG}$, $E^-_\mathrm{DDECPOS}$, and $E^-_\mathrm{QDECPOS}$, respectively, shown here in red. This operation results in voltages in two reference frames decoupled from the opposite reference frame, $E^+_\mathrm{DDEC}$, $E^+_\mathrm{QDEC}$, $E^-_\mathrm{DDEC}$, and $E^-_\mathrm{QDEC}$, shown in green. These decoupled voltages are filtered to obtain the cyan curves $E^+_\mathrm{DFILT}$, $E^+_\mathrm{QFILT}$, $E^-_\mathrm{DFILT}$, and $E^-_\mathrm{QFILT}$, and transformed into the opposite rotating reference frame to produce again the red curves $E^+_\mathrm{DDECNEG}$, $E^+_\mathrm{QDECNEG}$, $E^-_\mathrm{DDECPOS}$, and $E^-_\mathrm{QDECPOS}$, which are used to subtract the coupling from the input voltages. The filter is necessary in order to avoid the infinite-gain loop inherent in the decoupling system. The implementation in a digital controller requires a buffer for feeding back the filtered decoupled values.

\begin{figure}[!ht]
\centering
	\subfloat[][]{\includegraphics[width=0.35\linewidth]{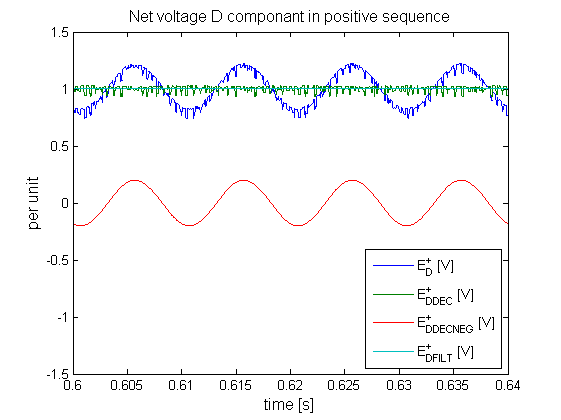}\label{g.Dcomp_plus}}
	\subfloat[][]{\includegraphics[width=0.35\linewidth]{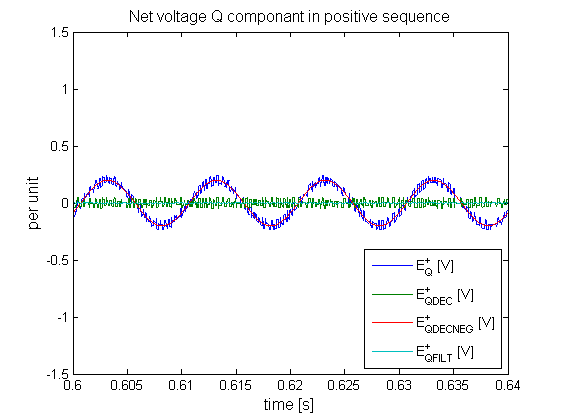}\label{g.Qcomp_plus}}\\
	\subfloat[][]{\includegraphics[width=0.35\linewidth]{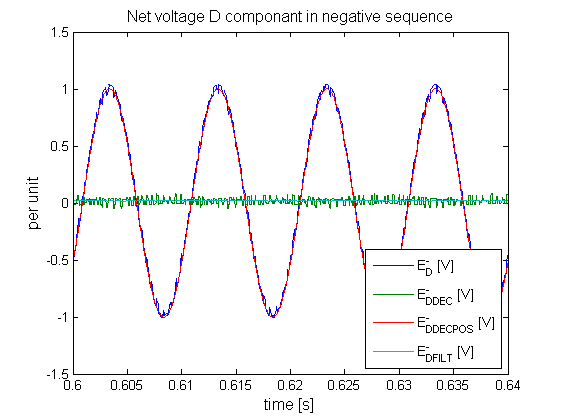}\label{g.Dcomp_neg}}
	\subfloat[][]{\includegraphics[width=0.35\linewidth]{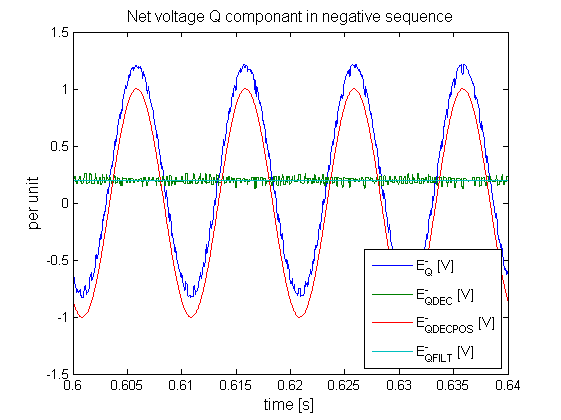}\label{g.Qcomp_neg}}
	\caption{Decoupling illustrated with values of (a) the positive-sequence D components, (b) positive-sequence Q components, (c) negative-sequence D components, and (d) negative-sequence Q components.}
	\label{g.Decoupling_illustrated}
\end{figure}

\begin{figure}[!ht]
\centering
	\subfloat[][]{\includegraphics[width=0.35\linewidth]{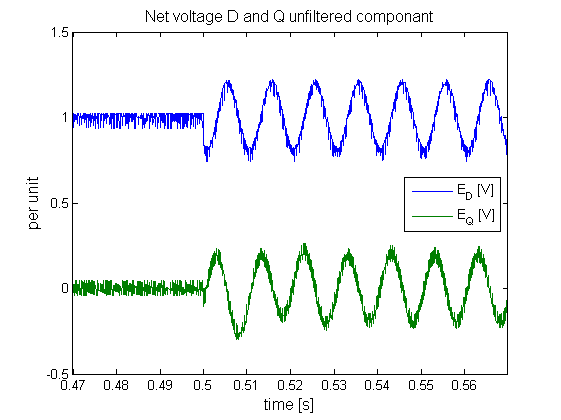}\label{g.NetDQ_comp_in}}
	\subfloat[][]{\includegraphics[width=0.35\linewidth]{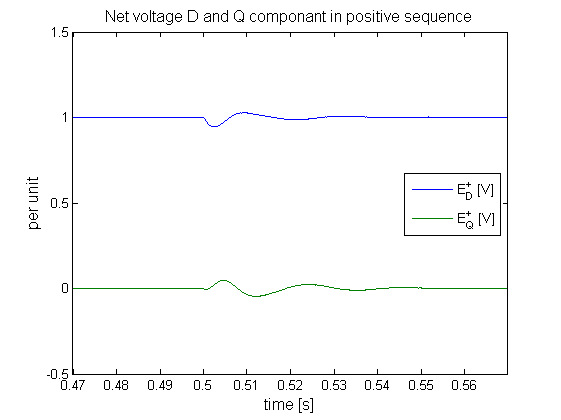}\label{g.NetDQ_comp_pos}}\\
	\subfloat[][]{\includegraphics[width=0.35\linewidth]{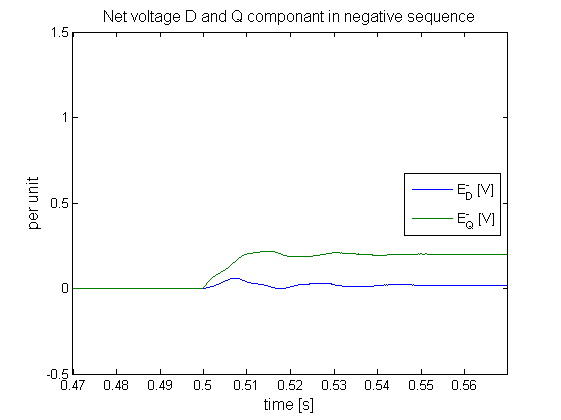}\label{g.NetDQ_comp_neg}}
	\subfloat[][]{\includegraphics[width=0.35\linewidth]{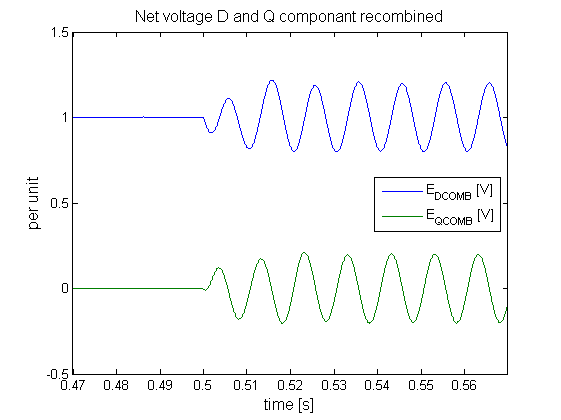}\label{g.NetDQ_comp_comb}}
	\caption{(a) Decoupling of the second-harmonic component in positive-sequence reference frame, (b) decoupled positive sequence, (c) decoupled negative sequence, and (d) recombination of decoupled sequences.}
	\label{g.DQDecoupled}
\end{figure}

When an imbalance occurs, as in the DQ representation of the voltages in
\Fref{g.DQDecoupled}\subref{g.NetDQ_comp_in}, the decoupling of the two opposite sequences is computed using a time transient due to the filtering of the feedback signal. The results of recombination of the two filtered and decoupled signals (Figs. \ref{g.DQDecoupled}\subref{g.NetDQ_comp_pos} and \ref{g.DQDecoupled}\subref{g.NetDQ_comp_neg}) contain only the fundamental frequency from the positive-sequence rotating frame and the second harmonic from the negative-sequence rotating frame. As the filtered signal provides a clean description of the second harmonic with no phase errors (\Fref{g.DQDecoupled}\subref{g.NetDQ_comp_comb}), this filtered signal can be used as an accurate feedforward value at the output of the current controller.

\section{Current control}
\label{sec:posnegcontrol}

The current controller for the voltage source inverters that was used in this work was similar to the classic PI controller operating in the synchronous rotating reference frame, but had superior decoupling characteristics \cite{intro.multiPI1,intro.multiPI2}. Dimensioning rules will be given for the optimal parameters as a function of the bandwidth of the whole system. For grid-connected converters operating at lower switching frequencies, the bandwidth does not allow one to compensate the second harmonic due to phase voltage imbalance. In the present work, an approach similar to that used for the decoupling of the second-harmonic component of the network voltages was adopted for the measured phase currents.
 In the positive sequence, the current vector is given by $I^+_\mathrm{D}$ and $I^+_\mathrm{Q}$, and the current vector in the negative sequence is given by $I^-_\mathrm{D}$ and $I^-_\mathrm{Q}$. If the currents are well balanced, the negative components $I^-_\mathrm{D}$ and $I^-_\mathrm{Q}$ are zero; if the currents are in phase with the voltages, the positive component $I^+_\mathrm{Q}$ is zero. Feeding the decoupled voltages in the positive and negative sequences forward helps the controller to compensate the current imbalance \cite{double.siemaszko}. However, for full control of the current (im)balance, we suggest a double-frame control approach, with one frame controller for each of the two sequences. This method is well known in the literature \cite{double.rufer,double.alepuz,double.czech,double.roiu,double.suh}; however, the proper decoupling of the sequences in the two rotating frames was introduced only in Ref. \cite{sequence.epe}, and details were given in Ref. \cite{double.teodorescu}. Other control methods exist in the stationary reference frame \cite{compensation.statcom}, but will not be considered here.

\subsection{Multivariable current controller for voltage source inverters}
\label{sub:multivariable}

The plant is described in the stationary reference frame as a function of the voltage phasor $\underline{U}$, the current phasor $\underline{I}$, and the passive elements $R_\sigma$ and $L_\sigma$ as in Eq. (\ref{equ.u}). The transformation to the rotating reference frame causes the voltage $\underline{U}_\mathrm{S}$ to appear as a component $j\omega L_\sigma \underline{I}_\mathrm{S}$ as in Eq. (\ref{equ.us}), which is the inherent coupling between the direct and quadrature components. In the classic PI controller, this component is compensated by a proportional term $\omega L_\sigma$, but mathematical analysis shows that the full compensation contains a cross-coupling with integrators, as demonstrated in Refs. \cite{intro.multiPI1,intro.multiPI2}, when the principle of the multivariable PI controller is applied:
\begin{eqnarray}
\underline{U}		& = & R_\sigma \underline{I} + L_\sigma \frac{\mathrm{d}(\underline{I})}{\mathrm{d}t}, \label{equ.u}\\
\underline{U}_\mathrm{S}		& = & R_\sigma \underline{I}e^{j\omega t} + L_\sigma \frac{\mathrm{d}(\underline{I}e^{j\omega t})}{\mathrm{d}t} = R_\sigma \underline{I}_\mathrm{S} + L_\sigma \frac{\mathrm{d}\underline{I}_\mathrm{S}}{\mathrm{d}t} + j\omega L_\sigma \underline{I}_\mathrm{S} = [R_\sigma + (s+j\omega)L_\sigma]\underline{I}_\mathrm{S}. \label{equ.us}
\end{eqnarray}

The feedback control principle is illustrated in Fig.~10(a). The block elements of the system, considered in the Laplace domain, describe the plant $G_\mathrm{S}(s)$, the modulator $G_\mathrm{M}(s)$, and the measurement of the current. The controller $G_\mathrm{R}(s)$ was defined in order to compensate the plant time constant $T_\mathrm{N}$. The integral time constant $T_\mathrm{I}$ was defined using the magnitude optimum criterion as given by Eq. (\ref{equ.system}):
\begin{equation}
\begin{cases}
G_\mathrm{S}(s)	 =	\cfrac{K_S}{1+(s+j\omega)T_\mathrm{S}}~, ~ G_\mathrm{M}(s)	 =  \cfrac{K_\mathrm{M}}{1+sT_\mathrm{M}}, \\
G_\mathrm{R}(s)	 =  \cfrac{1+(s+j\omega)T_\mathrm{N}}{sT_\mathrm{I}} = \cfrac{1+sT_\mathrm{N}}{sT_\mathrm{I}} + \cfrac{j\omega T_\mathrm{N}}{sT_\mathrm{I}},
\end{cases}
\begin{cases}
T_\mathrm{N}	 =	 T_\mathrm{S}, \\
T_\mathrm{I}	 =	 2K_\mathrm{S}K_\mathrm{M}T_\mathrm{M},
\end{cases}
\begin{cases}
T_\mathrm{S}	 =	 \cfrac{L_\sigma}{R_\sigma}, \\
K_\mathrm{S}	 =	 \cfrac{1}{R_\sigma},
\end{cases}\label{equ.system}
\end{equation}
\begin{equation}
G_0(s) 	 =  G_\mathrm{R}(s)G_\mathrm{M}(s)G_\mathrm{S}(s) = \frac{1 + (s+j\omega)T_\mathrm{N}}{sT_\mathrm{I}}\frac{K_\mathrm{M}}{1 + sT_\mathrm{M}}\frac{K_\mathrm{S}}{1 + (s+j\omega)T_\mathrm{S}} = \frac{1}{s2T_\mathrm{M}}\frac{1}{1 + sT_\mathrm{M}}. \label{equ.openloop}
\end{equation}
The open-loop transfer function $G_0(s)$ of the whole system is given by Eq. (\ref{equ.openloop}) and is illustrated in Fig.~10(b). The controller time constants were set in such a way that the open-loop transfer function crossed the zero axis with a slope of $-$20~dB/dec and a phase margin of $63^\circ$ was kept, ensuring the stability of the system. The bandwidth of the system was limited by the time constant of the modulator.

\begin{figure*}[!ht]
		\centering
		\subfloat[][]{\includegraphics[scale=0.5]{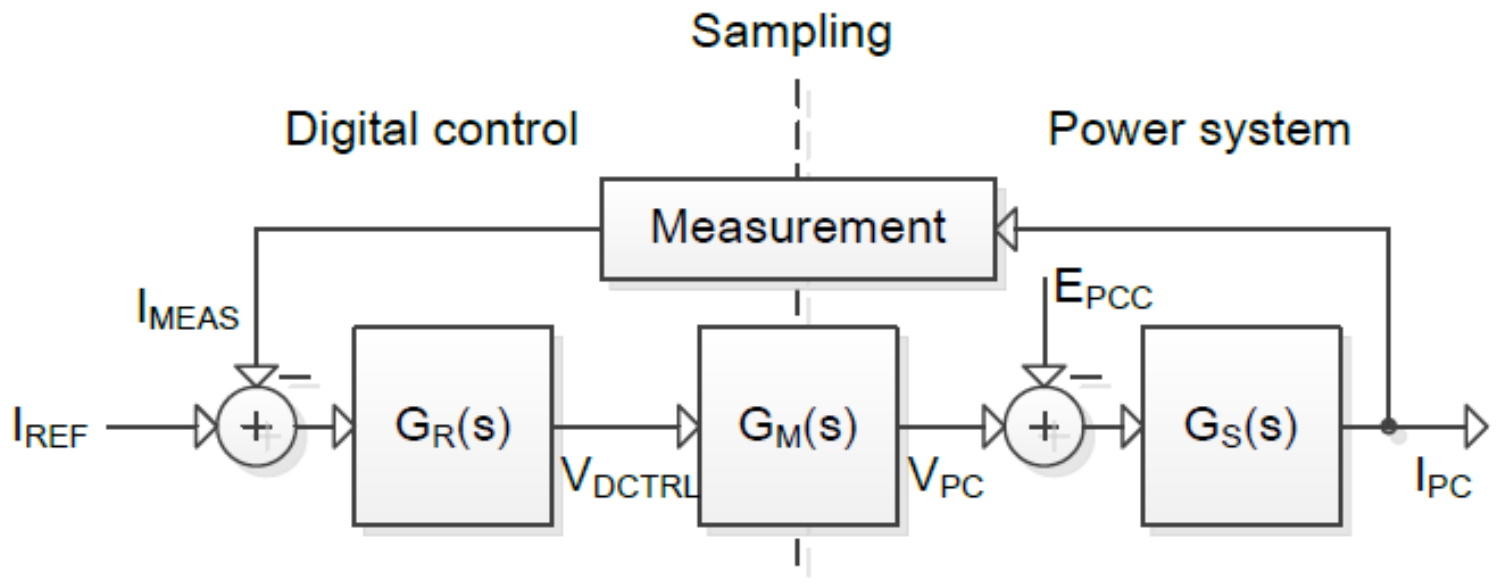}\label{g.controlprinciple}}
		\subfloat[][]{\includegraphics[scale=0.5]{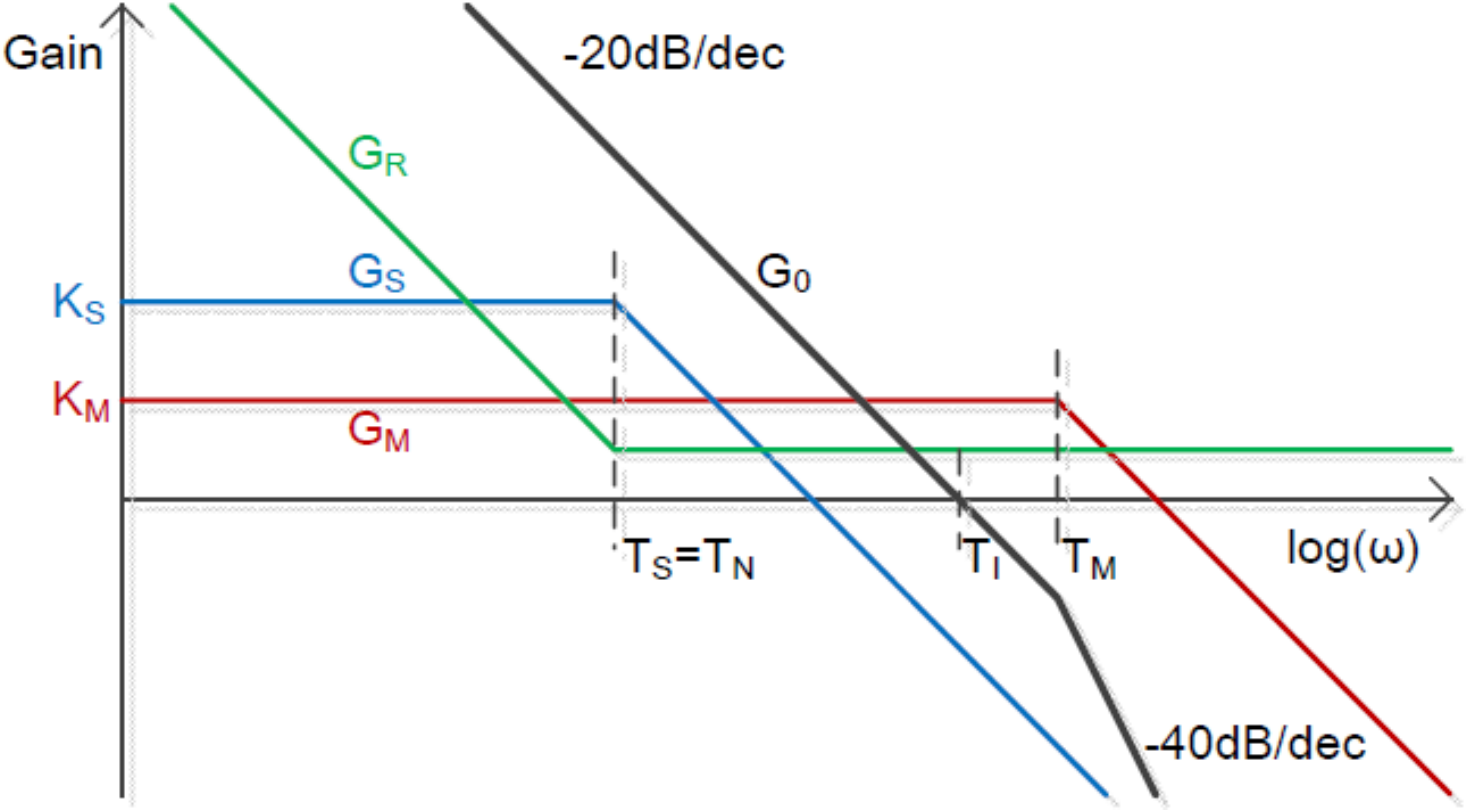}\label{g.CTRL_bandwidth}}
    \caption{(a) Principle of feedback control; (b) system bandwidth}
\end{figure*}

Finally, the control principle of the multivariable PI controller, leading to the control equation described in Eq. (\ref{equ.vdvq}), is illustrated in
\Fref{g.PIprinciple}\subref{g.PIprinciple_A}. The sampling of the measurement and the control is synchronized with the triangle of the pulse width modulation (PWM), which is synchronized with the PLL as illustrated in \Fref{g.PIprinciple}\subref{g.PIprinciple_B}, as  described in Ref. \cite{sampling.epe}. In such a way, one can operate the power converter at low frequencies by ensuring symmetry in the current signal. Moreover, the current measurement is performed  between two switching events, where an average value between two peaks is considered naturally and additional filters adding unnecessary phase shifts are not required:
\begin{equation}
\begin{cases}
V_\mathrm{D}		 =  \cfrac{1+sT_\mathrm{N}}{sT_\mathrm{I}}I_\mathrm{Derr} + \cfrac{\omega T_\mathrm{N}}{sT_\mathrm{I}}I_\mathrm{Qerr}, \\
V_\mathrm{Q} 		 = \cfrac{1+sT_\mathrm{N}}{sT_\mathrm{I}}I_\mathrm{Qerr} + \cfrac{\omega T_\mathrm{N}}{sT_\mathrm{I}}I_\mathrm{Derr}. \label{equ.vdvq}
\end{cases}
\end{equation}

\begin{figure}[!ht]
		\centering
		\subfloat[][]{\includegraphics[scale=0.5]{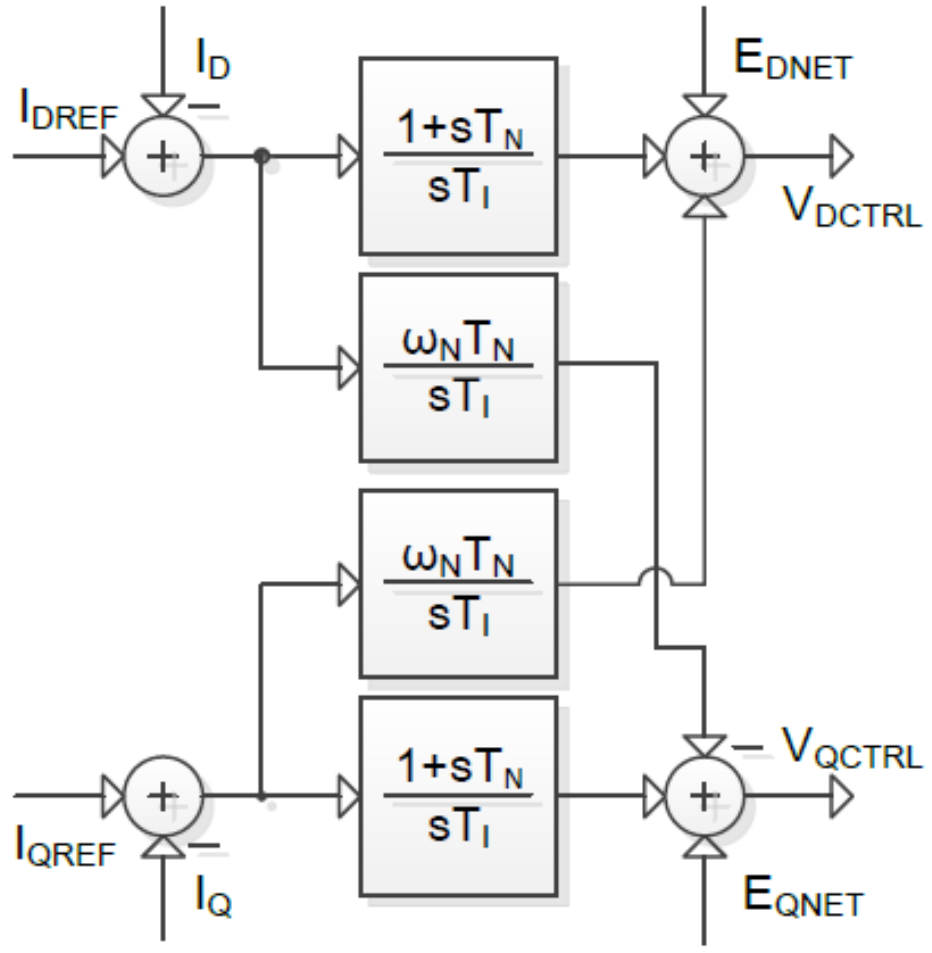}\label{g.PIprinciple_A}}
		\subfloat[][]{\includegraphics[scale=0.5]{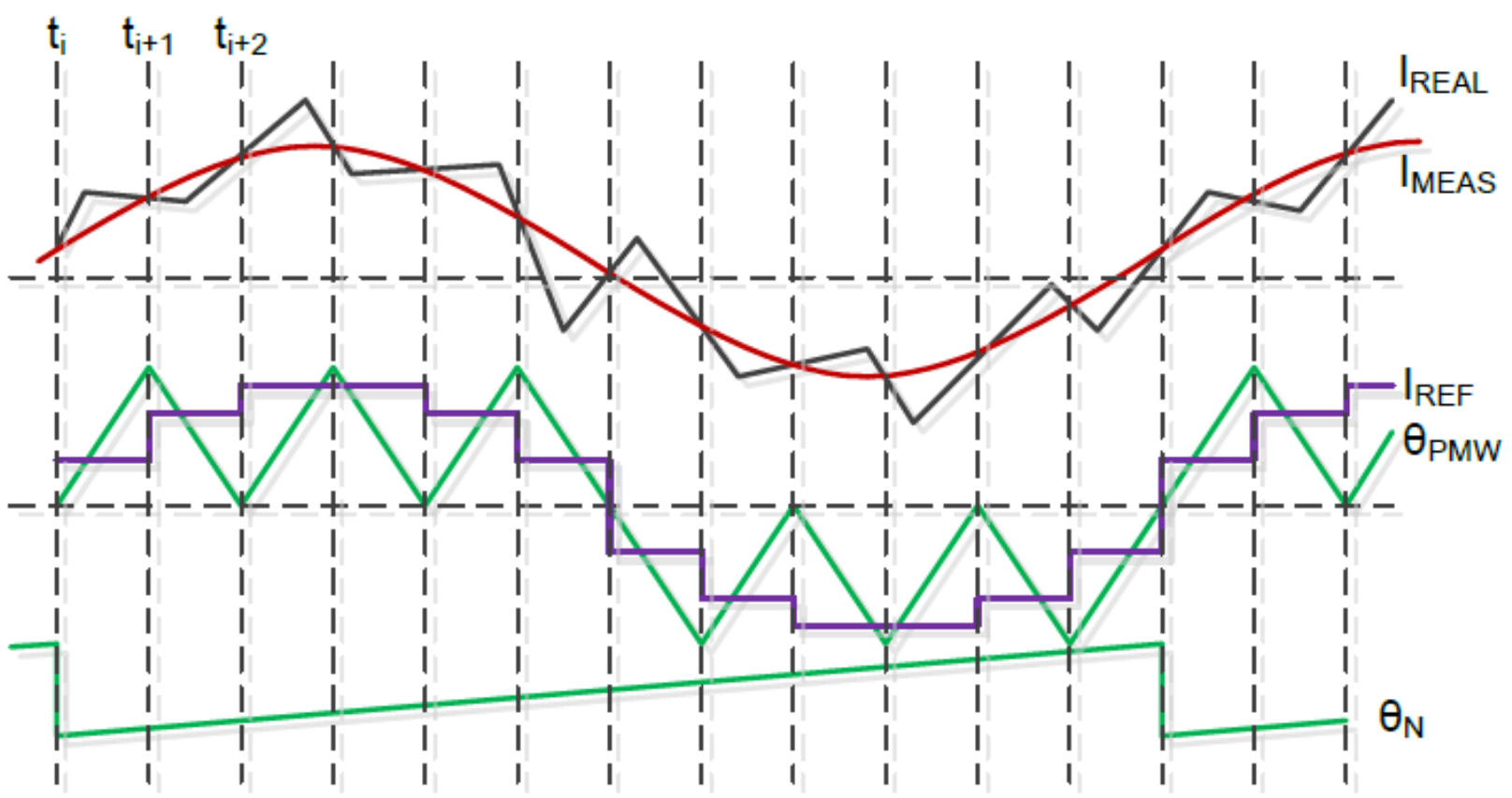}\label{g.PIprinciple_B}}
    \caption{Principle of (a) the multivariable PI controller and (b) sampling/modulation}
		\label{g.PIprinciple}
\end{figure}

\subsection{Control of the current in an asymmetric grid}
\label{sub:control_asym}

An approach similar to the decoupling of network voltages presented in the previous section was adopted to obtain four decoupled components from the measured currents. The current controller presented in this paper was implemented in the HIL simulation model. The control platform had its own individual sampling system, but the controller sampling and the pulse width modulator were synchronized with the PLL of the network, as illustrated in \Fref{g.PIprinciple}\subref{g.PIprinciple_B}, for all switching frequencies tested in the work described in this section. The second-harmonic component appearing in the voltages can be compensated by the current controller only if the bandwidth of the system allows it. As shown in the previous section, the control loop bandwidth depends mainly on the modulator, and specifically on the converter switching frequency.

To illustrate the bandwidth limitations of the controller, the power converter was operated at 450~Hz. The current controller was tested with two current steps under conditions of phase voltage imbalance. As illustrated in \Fref{g.single_frame}\subref{g.NetDQ_single_36}, the measurement of the four voltage components was not affected by the low switching frequency. One can see in \Fref{g.single_frame}\subref{g.CurDQ_pos_single_36_cross} the accuracy of the decoupling between the direct and the quadrature components performed by the multivariable PI controller. Also, the currents do not seem to be affected by the transient that appears when the voltage imbalance occurs. However, \Fref{g.single_frame}\subref{g.CurDQ_neg_single_36_cross} shows a negative sequence appearing in the currents, indicating that current balance is not ensured owing to a lack of bandwidth on the controller side.

\begin{figure}[!ht]
\centering
	\subfloat[][]{\includegraphics[width=0.34\linewidth]{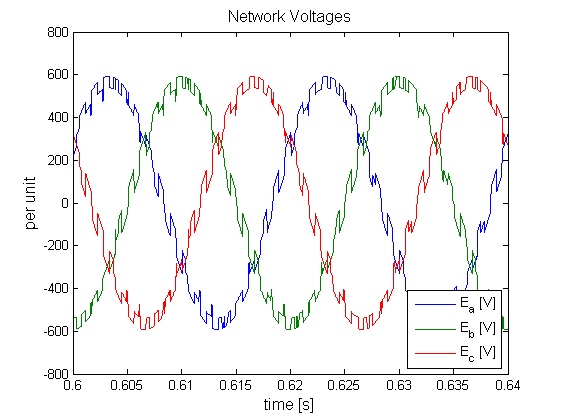}\label{g.Network_votlage_unbal_36}}
	\subfloat[][]{\includegraphics[width=0.34\linewidth]{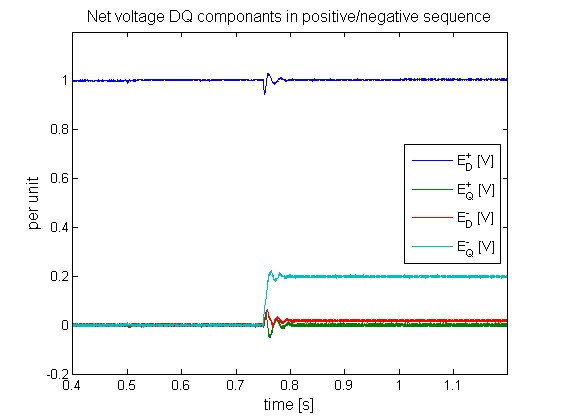}\label{g.NetDQ_single_36}}\\
	\subfloat[][]{\includegraphics[width=0.34\linewidth]{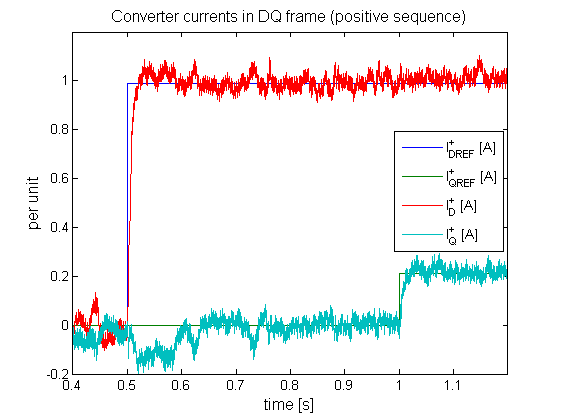}\label{g.CurDQ_pos_single_36_cross}}
	\subfloat[][]{\includegraphics[width=0.34\linewidth]{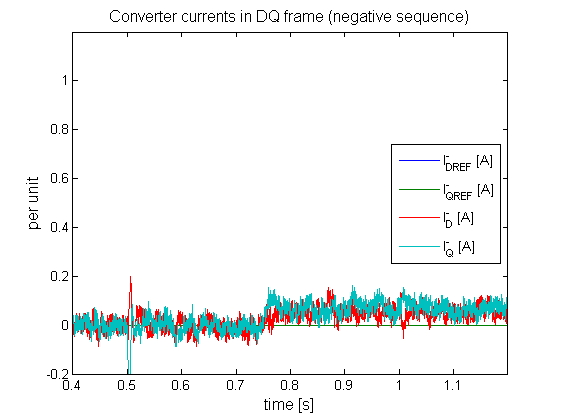}\label{g.CurDQ_neg_single_36_cross}}
	\caption{Operation at 450~Hz switching frequency: (a) phase voltages under balanced conditions, (b) decoupled voltage components, (c) currents and references in the positive sequence, and (d) the negative sequence.}
	\label{g.single_frame}
\end{figure}

\subsection{Balance control of the three phase currents under asymmetric phase voltage conditions}
\label{sub:ddsrf_cc}

In order to accurately control each of the four current components distributed in the two rotating reference frames, two current controllers are needed, one for each rotating frame, with a proper decoupling of the four components. The \emph{double-decoupled synchronous-reference-frame multivariable PI current controller}, illustrated in Fig. \ref{g.currentcontrol}, consists of two single-frame current controllers implemented in mirror form for both the positive and the negative sequence.

\begin{figure}[!ht]
		\centering
    \includegraphics[scale=0.48]{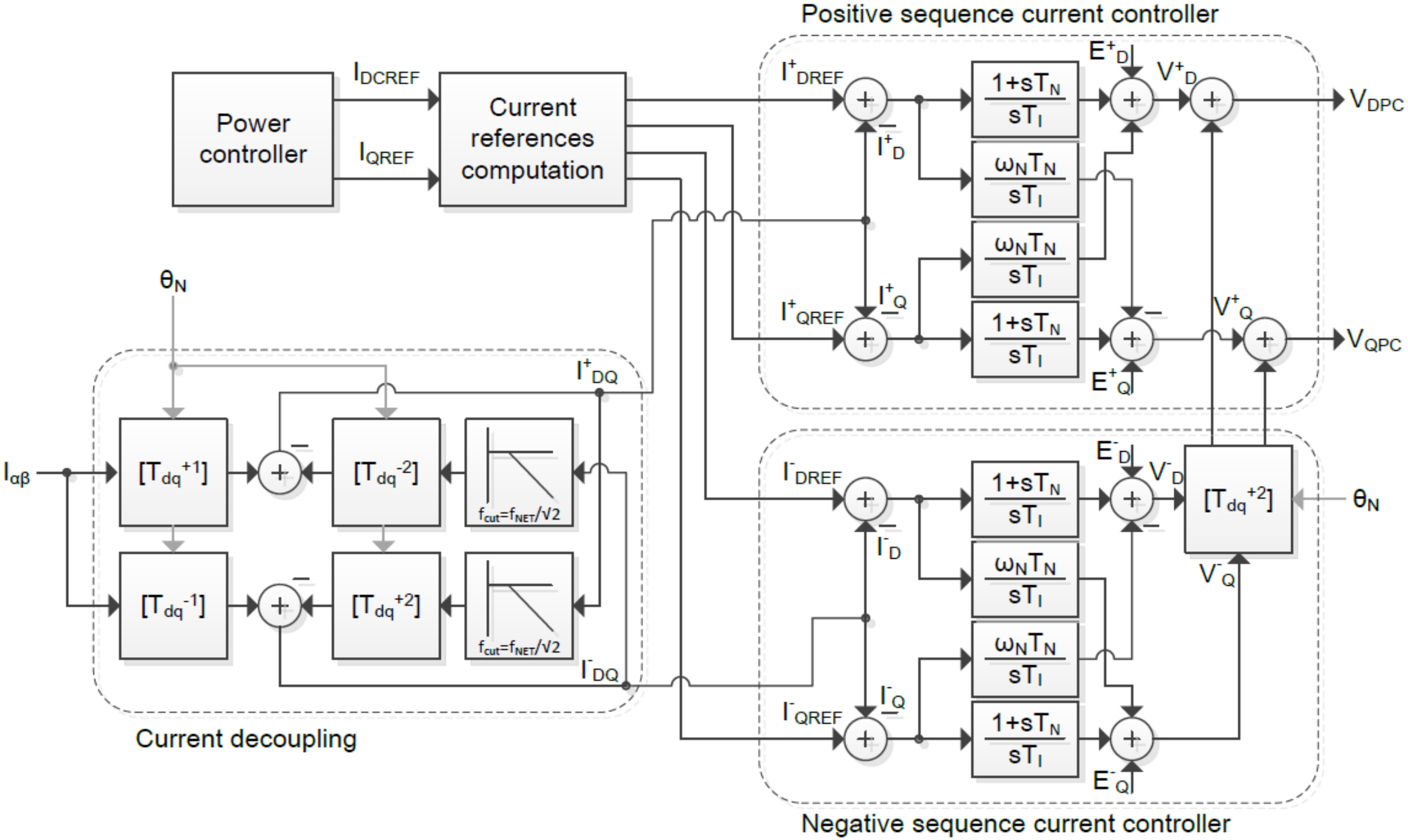}
    \caption{Double-decoupled synchronous-reference-frame multivariable current controller}
    \label{g.currentcontrol}
\end{figure}

In the positive reference frame, the direct and quadrature current references $I^+_\mathrm{DREF}$ and $I^+_\mathrm{QREF}$ are applied to the two cross-coupled PI controllers regulating the network currents $I^+_\mathrm{D}$ and $I^+_\mathrm{Q}$ before the decoupled network voltages $E^+_\mathrm{D}$ and $E^+_\mathrm{Q}$ are applied as a feedforward signal. In the negative frame, the same scheme is applied in mirror image form, with inverted signs in the cross-coupling contributions. The output voltage reference signals $V^-_\mathrm{D}$ and $V^-_\mathrm{Q}$ are transformed to the positive sequence before being added to $V^+_\mathrm{D}$ and $V^+_\mathrm{Q}$ as a contribution for the unbalanced phase voltages before being applied by the modulator.

The current references $I^-_\mathrm{DREF}$ and $I^-_\mathrm{QREF}$ in the negative sequence can be set to zero to ensure current symmetry in the case of imbalance in the phase voltages. With values different from zero, the current references $I^-_\mathrm{DREF}$ and $I^-_\mathrm{QREF}$ can be used to control the negative-sequence currents in order to introduce imbalance into the phase currents.

The double-frame controller was tested under conditions of voltage imbalance with four current reference steps.
Figure \ref{g.double_frame}\subref{g.CurDQ_pos_double_120}
shows that the currents in the positive sequence were well controlled and decoupled. In the negative sequence, as illustrated in
\Fref{g.double_frame}\subref{g.CurDQ_neg_double_120},
current symmetry could be ensured by control; at the same time, we could control the imbalance freely to obtain currents as in
\Fref{g.double_frame}\subref{g.Converter_currents_double_120}.
After the decoupling between the direct and the quadrature components of the currents by the multivariable PI approach had been verified,
\Fref{g.double_frame}\subref{g.CurDQ_pos_double_120}
shows that a small transient appeared in the negative sequence when a current step was applied in the positive sequence. The reason comes from the necessary filtering applied to the feedback signals and the fact that the sudden phase imbalance appears as a step in time.

\begin{figure*}[!ht]
\centering
	\subfloat[][]{\includegraphics[width=0.36\linewidth]{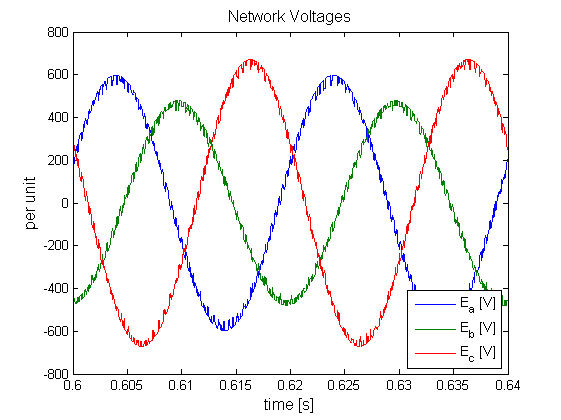}\label{g.Network_votlage_unbal}}
	\subfloat[][]{\includegraphics[width=0.36\linewidth]{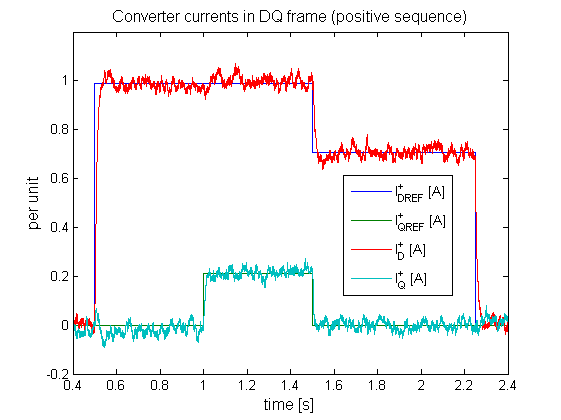}\label{g.CurDQ_pos_double_120}}\\
	\subfloat[][]{\includegraphics[width=0.36\linewidth]{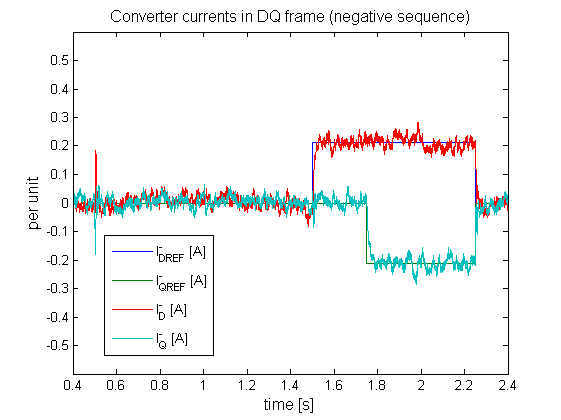}\label{g.CurDQ_neg_double_120}}
	\subfloat[][]{\includegraphics[width=0.36\linewidth]{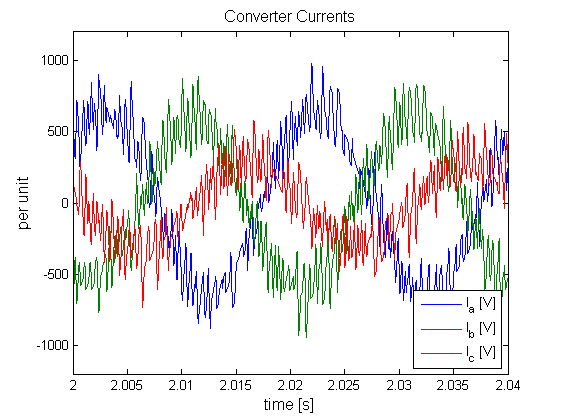}\label{g.Converter_currents_double_120}}
	\caption{Double-frame current control: (a) phase voltage imbalance, (b) current control in the positive sequence, (c) current control in the negative sequence, and (d) resulting phase currents.}
	\label{g.double_frame}
\end{figure*}

\section{Power considerations}
\label{sec:powerCompensation}

The previous section introduced a simple way to control the four components of the phase current, independently of the state of balance of the phase voltages. When imbalance occurs, the second harmonic appearing in the phase voltages affects the instantaneous power seen at the point of common coupling. If uncontrolled, the phase currents tend to have an imbalance in the same direction as the imbalance in the phase voltages. This strongly affects the instantaneous power as seen at the point of common coupling and also on the DC link side, as pointed out in Ref. \cite{DC.epe}. When a phase voltage imbalance occurs and the controller maintains symmetry in the phase currents, the second harmonic in the instantaneous power is considerably reduced, and the ripple in the DC link voltage is reduced at the same time. Another way is to act on the current references so as to impose an imbalance in the phase currents to counter the phase voltage imbalance. For this compensation to be implemented, the instantaneous power must be computed as a function of the four current and four voltage components to assess the current references as a function of the voltage imbalance.

\subsection{Power oscillations during network imbalance}
\label{sub:poweroscillation}

When an asymmetric perturbation occurs in the phase voltages, as in
\Fref{g.power_computations}\subref{g.m0_Voltages_sscc8_sf180},
and if symmetric current operation is maintained as in
\Fref{g.power_computations}\subref{g.m0_Currents_sscc8_sf180}, a second harmonic appears in the computation of the active and reactive instantaneous power as in
\Fref{g.power_computations}\subref{g.m0_Power_sscc8_sf180}. This power oscillation is even stronger if the negative sequence of the current is not controlled, since the current imbalance tends to be in the same direction as the voltage imbalance (as in
Figs. \ref{g.single_frame}\subref{g.NetDQ_single_36} and
\ref{g.single_frame}\subref{g.CurDQ_neg_single_36_cross}). A simple relationship established in Ref. \cite{theory.yazdani} links the power ripple to the DC link ripple, as shown in Eq. (\ref{equ.ripple}):
\begin{equation}
V_\mathrm{DCripple}	 = \frac{P_\mathrm{ripple}}{2C_\mathrm{DC}\omega V_\mathrm{DC}}. \label{equ.ripple}
\end{equation}
There is an evident relationship between the voltage ripple and the value of the DC link capacitance, which in industrial applications tends to be as low as possible. Therefore, to maintain the DC voltage ripple at its minimum, one must act on the power ripple either by operating with symmetric currents or by compensating the phase voltage imbalance with an opposite phase current imbalance.

\begin{figure*}[!ht]
\centering
	\subfloat[][]{\includegraphics[width=0.36\linewidth]{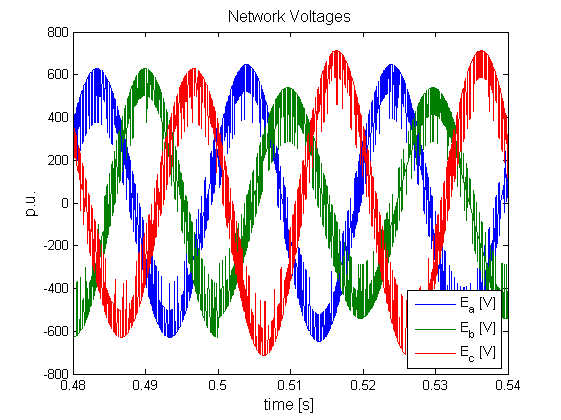}\label{g.m0_Voltages_sscc8_sf180}}
	\subfloat[][]{\includegraphics[width=0.36\linewidth]{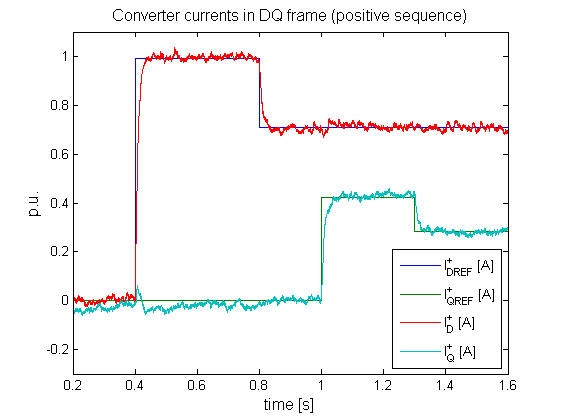}\label{g.m0_DQpositive_sscc8_sf180}}\\
	\subfloat[][]{\includegraphics[width=0.36\linewidth]{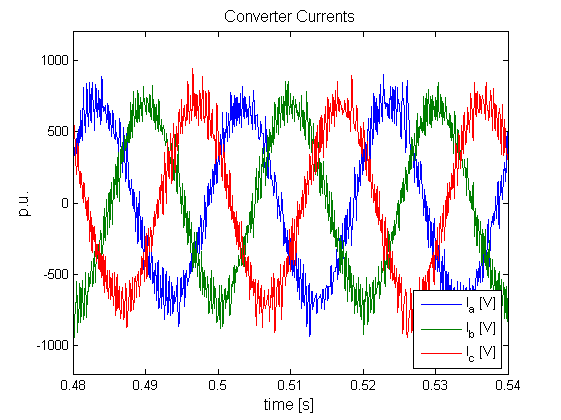}\label{g.m0_Currents_sscc8_sf180}}
	\subfloat[][]{\includegraphics[width=0.36\linewidth]{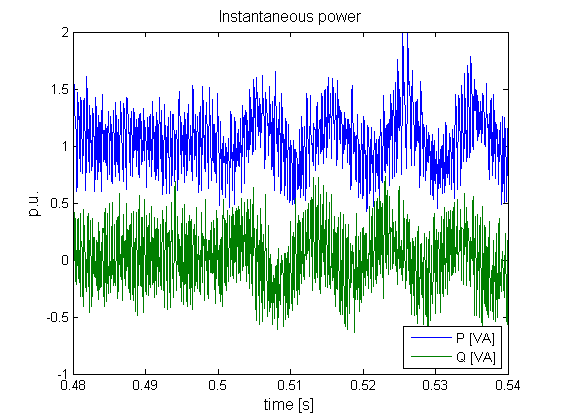}\label{g.m0_Power_sscc8_sf180}}
	\caption{Power considerations for unbalanced network: (a) phase voltages, (b) control of positive-sequence currents, (c) resulting phase currents, and (d) instantaneous active and reactive power.}
	\label{g.power_computations}
\end{figure*}

\subsection{Elements of instantaneous-power theory}
\label{sub:powertheory}

One approach to defining the reactive instantaneous power was introduced in Japan in the 1980s and has been considered as a reference \cite{theory.Akagi,theory.furuhushi,theory.nabae}. In contrast to the German approach known as the FBD method \cite{theory.fbd}, this instantaneous-power theory has led to many further developments and attempts to find a unified theory. Some other approaches that have introduced different definitions of power components, such as Refs. \cite{power.epe,power2.epe}, have shown the state of uncertainty and lack of agreement in the scientific community \cite{theory.kim,theory.Czarnecky,theory.Herrara,theory.tenti}. In the Japanese approach, the apparent power is a combination of continuous parts $P_\mathrm{NET}$ and $Q_\mathrm{NET}$ and four redundant second-harmonic oscillating parts $P_\mathrm{2C}$, $P_\mathrm{2S}$, $Q_\mathrm{2C}$, and $Q_\mathrm{S}$, as in Eq. (\ref{equ.snetpow}). The active and reactive power are found by taking the real and imaginary parts, respectively, of the apparent complex power $S_\mathrm{NET}$:
\begin{equation}
S_\mathrm{NET} =	(P + P_\mathrm{2C}\cos(2\omega t) + P_\mathrm{2S}\sin(2\omega t)) + j(Q + Q_\mathrm{2C}\cos(2\omega t) + Q_\mathrm{2S}\sin(2\omega t)). \label{equ.snetpow}
\end{equation}

The concept of the quadrature complex power $T_\mathrm{NET}$ was introduced in Ref. \cite{theory.suh}. The reactive power is calculated on the basis of the conjugate current vector $I_\mathrm{DQS}^\ast$ and a voltage vector $E_\mathrm{DQS}'$ lagging the pole voltage vector $E_\mathrm{DQS}$ by $90^\circ$. Following this approach, the active power $P_\mathrm{NET}$ is the real part of the apparent complex power $S_\mathrm{NET}$ and the reactive power $Q_\mathrm{NET}$ is the real part of the quadrature complex power $T_\mathrm{NET}$ as in Eq. (\ref{equ.pnet}):
\begin{equation}
\begin{cases}
S_\mathrm{NET} 			  = 	\frac{3}{2}E_\mathrm{DQS}I_\mathrm{DQS}^\ast ,\\
T_\mathrm{NET} 			  = 	\frac{2}{3}E_\mathrm{DQS}'I_\mathrm{DQS}^\ast ,
\end{cases}
\begin{cases}
I_\mathrm{DQS}^\ast 	 =  e^{j\omega_\mathrm{N} t}I^+_\mathrm{DQ} - e^{-j\omega_\mathrm{N} t}I^-_\mathrm{DQ} , \\
E_\mathrm{DQS} 			 =  e^{j\omega_\mathrm{N} t}E^+_\mathrm{DQ} + e^{-j\omega_\mathrm{N} t}E^-_\mathrm{DQ} , \\
E_\mathrm{DQS}' 			 =  -je^{j\omega_N t}E^+_\mathrm{DQ} + je^{-j\omega_\mathrm{N} t}E^-_\mathrm{DQ} ,
\end{cases}
\begin{cases}
P_\mathrm{NET}				 =  \operatorname{Re}\left\{S_\mathrm{NET}\right\} , \\
Q_\mathrm{NET}				 =  \operatorname{Re}\left\{T_\mathrm{NET}\right\} . \label{equ.pnet}
\end{cases}
\end{equation}
In any case, the instantaneous power components can be computed as in Eq. (\ref{equ.power}) and can serve as a basis for the computation of the compensation, in a way similar to that proposed in Ref. \cite{compensation.epe}:
\begin{equation}
\left[ \begin{array}{c}
P \\ Q \\ P_\mathrm{2C} \\ P_\mathrm{2S}
\end{array}\right] = \frac{3}{2}
\left[ \begin{array}{rrrr}
E^+_\mathrm{D} & E^+_\mathrm{Q} & E^-_\mathrm{D} & E^-_\mathrm{Q} \\ E^+_\mathrm{Q} & -E^+_\mathrm{D} & -E^-_\mathrm{Q} & E^-_\mathrm{D} \\ E^-_\mathrm{D} & E^-_\mathrm{Q} & E^+_\mathrm{D} & E^+_\mathrm{Q} \\ E^-_\mathrm{Q} & -E^-_\mathrm{D} & -E^+_\mathrm{Q} & E^+_\mathrm{D}
\end{array}\right]
\left[ \begin{array}{r}
I^+_\mathrm{D} \\ I^+_\mathrm{Q} \\ I^-_\mathrm{D} \\ I^-_\mathrm{Q}
\end{array}\right] .
\label{equ.power}
\end{equation}

\subsection{Compensation methods}
\label{sub:Song}

The relation presented in Eq. (\ref{equ.power}) is the starting point for establishing current references for reducing the second-harmonic content of the active power. By setting the oscillatory parts $P_\mathrm{2C}$ and $P_\mathrm{2S}$ of the complex power to zero, one can perform a simple matrix computation of the voltages to find the corresponding current references. This approach was introduced in Ref. \cite{power.song}. Another method was introduced in Ref. \cite{power.suh}, which considered the voltage drop across the filter impedance; in this reference, the original approach was criticized for containing a singularity. As demonstrated in Ref. \cite{double.siemaszko}, however, the second approach contained the same singularity. Finally, a third similar approach was introduced in Ref. \cite{power.siemaszko}, in which the strength of the network was considered. But, from the point of view of current control, none of those three methods can be distinguished from the others. Therefore, only the original method, described by Eq. (\ref{equ.song}), is presented here, for simplicity:
\begin{equation}
\left[ \begin{array}{r}
I^+_\mathrm{D} \\ I^+_\mathrm{Q} \\ I^-_\mathrm{D} \\ I^-_\mathrm{Q}
\end{array}\right] \approx
\left[ \begin{array}{rrrr}
E^+_\mathrm{D} & E^+_\mathrm{Q} & E^-_\mathrm{D} & E^-_\mathrm{Q} \\ E^+_\mathrm{Q} & -E^+_\mathrm{D} & -E^-_\mathrm{Q} & E^-_\mathrm{D} \\ E^-_\mathrm{D} & E^-_\mathrm{Q} & E^+_\mathrm{D} & E^+_\mathrm{Q} \\ E^-_\mathrm{Q} & -E^-_\mathrm{D} & -E^+_\mathrm{Q} & E^+_\mathrm{D}
\end{array}\right]^{-1}
\left[ \begin{array}{r}
P \\ Q \\ 0 \\ 0
\end{array}\right] .
\label{equ.song}
\end{equation}

The original method consists in solving the equations given in Eq. (\ref{equ.song}), where the current references are computed from the four components of the phase voltages as seen at the point of common coupling. The determinant $\mathrm{Det}$ of the condition matrix is calculated as in Eq. (\ref{sol.song}). A singularity appears when $D$ reaches zero, and therefore this method is limited and cannot compensate for a full phase loss.
\begin{equation}
\left[ \begin{array}{r}
I^+_\mathrm{DREF} \\ I^+_\mathrm{QREF} \\ I^-_\mathrm{DREF} \\ I^-_\mathrm{QREF}
\end{array}\right] \approx
\frac{P}{D}
\left[ \begin{array}{r}
E^+_\mathrm{D} \\ E^+_\mathrm{Q} \\ -E^-_\mathrm{D} \\ -E^-_\mathrm{Q}
\end{array}\right] +
\frac{Q}{D}
\left[ \begin{array}{r}
E^+_\mathrm{Q} \\ -E^+_\mathrm{D} \\ E^-_\mathrm{Q} \\ -E^-_\mathrm{D}
\end{array}\right]
\begin{cases}
\mathrm{Det}  = -D^2 , \\
D  =  \Big[(E^+_\mathrm{D})^2+(E^+_\mathrm{Q})^2\Big] - \Big[(E^-_\mathrm{D})^2+(E^-_\mathrm{Q})^2\Big] .
\end{cases}
\label{sol.song}
\end{equation}

A simulation of the power converter was run with the previously presented double-frame controller under several conditions of phase voltage balance and imbalance. The power compensation approach was implemented as a function of the four components of the phase voltages to compute the four current references. The system was run with several steps in the active and reactive currents, and the four current references were computed as a function of the four voltage components.

From the point of view of current control, all three compensation methods showed exactly the same results as for compensation of the active power at the point of common coupling, regardless of the strength of the network, the switching frequency, or the value of the line impedance on the power converter side. The compensation of the active-power oscillations could therefore be done by the method suggested in Ref. \cite{power.song}, which requires the lowest computation effort. When voltage imbalance appeared, the current references were all adjusted as in
Figs. \ref{g.song}\subref{g.m1_DQpositive_sscc8_sf180} and
\ref{g.song}\subref{g.m1_DQnegative_sscc8_sf180}. As predicted, the resulting phase currents, illustrated in
\Fref{g.song}\subref{g.Currents_sscc8_sf180_m1}, allowed us to compensate the active-power oscillations as in
\Fref{g.song}\subref{g.m1_Power_sscc8_sf180} by increasing the reactive-power oscillations.

\begin{figure*}[!ht]
\centering
	\subfloat[][]{\includegraphics[width=0.35\linewidth]{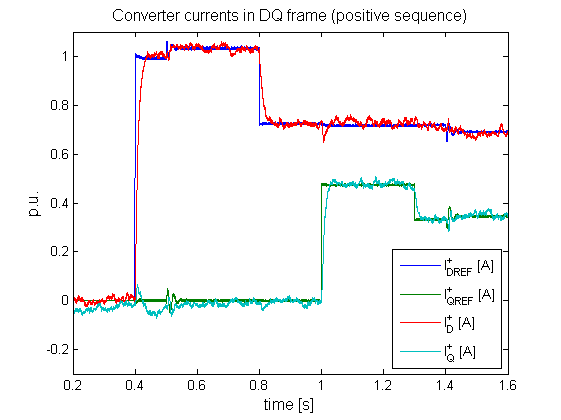}\label{g.m1_DQpositive_sscc8_sf180}}
	\subfloat[][]{\includegraphics[width=0.35\linewidth]{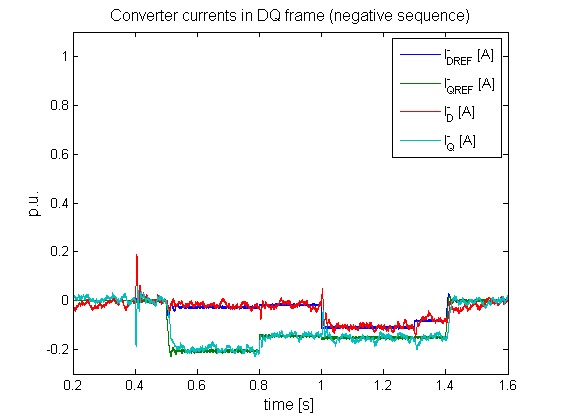}\label{g.m1_DQnegative_sscc8_sf180}}\\
	\subfloat[][]{\includegraphics[width=0.35\linewidth]{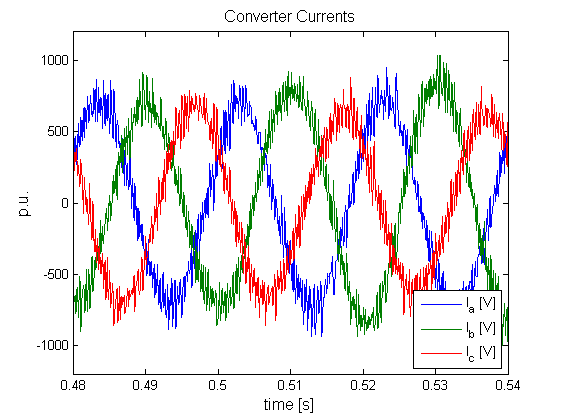}\label{g.Currents_sscc8_sf180_m1}}
	\subfloat[][]{\includegraphics[width=0.35\linewidth]{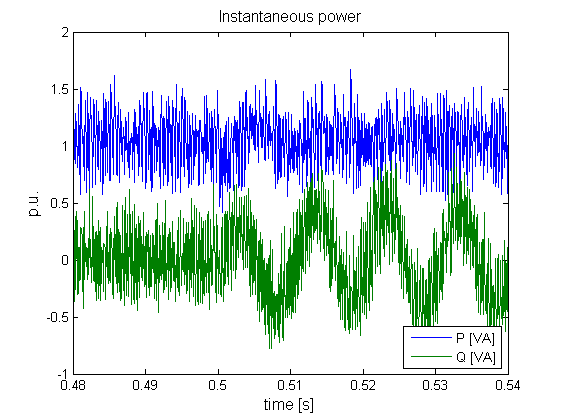}\label{g.m1_Power_sscc8_sf180}}
	\caption{(a) Active-power-oscillation compensation with current references in the positive sequence, (b) computed current references in the negative sequence, (c) resulting phase currents, and (d) instantaneous active and reactive power.}
	\label{g.song}
\end{figure*}

As long as the phase voltage dip is within some established limits, given by the condition that the positive-sequence components of the voltage remain higher than the negative-sequence components, one can increase the corresponding phase current to maintain the instantaneous active-power balance on the network side. Such operation is obviously limited by converter limits such as the maximum semiconductor device current, and the maximum phase voltage dip that can be handled by such a control method is dependent on the operating point of the whole power converter at the moment of the phase voltage dip.

\section{Conclusions}\label{sec:discussions}

When a voltage source inverter is operated with any type of grid, the synchronization must be robust enough to be able to reject all harmonics coming from the grid, as well as from the power converter itself. With respect to rejecting the second-harmonic component due to phase voltage imbalance and ensuring good PLL dynamics at the same time, the use of a decoupling method for the positive and negative sequences proved to be sufficiently accurate and allowed the current controller to work with four decoupled voltage components to handle asymmetries.

The current controller, when optimally tuned, showed its limitations in terms of bandwidth, especially at low switching frequencies. For the proper handling of currents under conditions of voltage asymmetry, feeding the decoupled values of the voltage forward is already a step forward. However, to ensure current symmetry, a second frame controller must be implemented to ensure zero values in the negative sequence of the currents.

One can control the current imbalance in such a way that the active-power oscillations due to voltage imbalance are reduced. A current reference computation method based on instantaneous-power theories has been presented and implemented. For phase dips down to a certain level, this compensation proved to work accurately, as long as the semiconductors allowed an increase in the current in some of the phases.


\begin{thebibliography}{99}


 \bibitem{pll.siemaszko}
D. Siemaszko, Double frame control and power compensation for power converters connected to weak networks with disturbances, Proc. 15th European Conf. on Power Electronics and Applications (EPE), 2013.

\bibitem{rl.cobreces}
S. Cobreces, E. Bueno, F.J. Rodriguez, F. Huerta, and P. Rodriguez, Influence analysis of the effects of an inductive-resistive weak grid over L and LCL filter current hysteresis controllers, Proc. 2007 European Conf. on Power Electronics and Applications, 2007.

\bibitem{pll.BenhabibEPE}
M.C. Benhabib, Fei Wang, and J.L. Duarte, Improved robust phase locked loop for utility grid applications, Proc. 13th European Conf. on Power Electronics and Applications, 2009 (EPE '09).

\bibitem{pll.RoblesEPE}
E. Robles, J. Pou, S. Ceballos, I. Gabiola, and M. Santos, Grid sequence detector based on a stationary reference frame, Proc. 13th European Conf. on Power Electronics and Applications, 2009 (EPE '09).

\bibitem{pll.suulEPE}
J.A. Suul, K. Ljokelsoy, and T. Undeland, Design, tuning and testing of a flexible PLL for grid synchronization of three-phase power converters, Proc. 13th European Conf. on Power Electronics and Applications, 2009 (EPE '09).

\bibitem{pll.rodriguezEPE}
P. Rodriguez, A. Luna, R. Teodorescu, F. Iov, and F. Blaabjerg, Fault ride-through capability implementation in wind turbine converters using a decoupled double synchronous reference frame PLL, Proc. 2007 European Conf. on Power Electronics and Applications, 2007.

\bibitem{pll.rodriguez}
 P. Rodriguez, J. Pou, J. Bergas, J.I. Candela, R.P. Burgos, and D. Boroyevich, \emph{IEEE Trans. Power Electron.} \textbf{22}(2) (2007) 584.

\bibitem{pll.carugati}
 I. Carugati, S. Maestri, P.G. Donato, D. Carrica, and M. Benedetti, \emph{IEEE Trans. Power Electron.} \textbf{27}(1) (2012) 321.

\bibitem{intro.multiPI1}
B. Bahrani, S. Kenzelmann, and A. Rufer, \emph{IEEE Trans. Ind. Electron.} \textbf{58}(7) (2011) 3016.
	
\bibitem{intro.multiPI2}
B. Bahrani, A. Karimi, B. Rey, and A. Rufer, \emph{IEEE Trans. Ind. Electron.} \textbf{60}(4) (2013) 1356.

\bibitem{double.siemaszko}
D. Siemaszko, Grid synchronization of power converters to weak unbalanced networks with disturbances, Proc. Electrical Systems for Aircraft, Railway and Ship Propulsion (ESARS), 2012.

\bibitem{double.rufer}
I. Etxeberria-Otadui, U. Viscarret, M. Caballero, A. Rufer, and S. Bacha, \emph{IEEE Trans. Ind. Electron.} \textbf{54}(5) (2007) 2902.
	\bibitem{double.alepuz}
S. Alepuz, S. Busquets-Monge, J. Bordonau, J.A. Martinez-Velasco, C.A. Silva, J. Pontt, and J. Rodriguez, \emph{IEEE Trans. Ind. Electron.} \textbf{56}(6) (2009) 2162.

\bibitem{double.czech}
Z.R. Ivanovic, E.M. Ad\v{z}i\'{c}, M.S. Veki\'{c}, S.U. Grabi\'{c}, N.L. \v{C}elanovi\'{c}, and V.A. Kati\'{c}, \emph{IEEE Trans. Power Electron.} \textbf{27}(11) (2012) 4699.

\bibitem{double.roiu}
D. Roiu, R.I. Bojoi, L.R. Limongi, and A. Tenconi, \emph{IEEE Trans. Ind. Appl.} \textbf{46}(1) (2010) 268.

\bibitem{double.suh}
Y. Suh, Y. Go, and D. Rho, \emph{IEEE Trans. Ind. Appl.} \textbf{47}(3) (2011) 1419.

\bibitem{sequence.epe}
S. Alepuz, S. Busquets, J. Bordonau, J. Pontt, C. Silva, and J. Rodriguez, Balanced grid currents in three-level voltage-source inverters connected to the utility under distorted condition using symmetrical components and linear quadratic regulator, Proc. 2007 European Conf. on Power Electronics and Applications, 2007.

\bibitem{double.teodorescu}
M. Reyes, P. Rodriguez, S. Vazquez, A. Luna, R. Teodorescu, and J.M. Carrasco, \emph{IEEE Trans. Power Electron.} \textbf{27}(9) (2012) 3934.

\bibitem{compensation.statcom}
I. Etxeberria-Otadui, U. Viscarret, I. Zamakona, B.A. Redondo, and J. Ibiricu, Improved STATCOM operation under transient disturbances for wind power applications, Proc. 2007 European Conf. on Power Electronics and Applications, 2007.

\bibitem{sampling.epe}
N. Hoffmann, F.W. Fuchs, and J. Dannehl, Models and effects of different updating and sampling concepts to the control of grid-connected PWM converters -- A study based on discrete time domain analysis, Proc. 14th European Conf. on Power Electronics and Applications (EPE 2011), 2011.

\bibitem{DC.epe}
J.G. Hwang, P.W. Lehn, and M. Winkelnkemper, Control of grid connected AC--DC converters with minimized DC link capacitance under unbalanced grid voltage condition, Proc. 2007 European Conf. on Power Electronics and Applications, 2007.

\bibitem{theory.yazdani}
A. Yazdani and R. Iravani, \emph{IEEE Trans. Power Deliv.} \textbf{21}(3) (2006) 1620.

\bibitem{theory.Akagi}
H. Akagi, Y. Kanazawa, and A. Nabae, \emph{IEEE Trans. Ind. Appl.} \textbf{IA-20}(3) (1984) 625.

\bibitem{theory.furuhushi}
T. Furuhashi, S. Okuma, and Y. Uchikawa, \emph{IEEE Trans. Ind. Electron.} \textbf{37}(1) (1990) 86.

\bibitem{theory.nabae}
A. Nabae and T. Tanaka, \emph{IEEE Trans. Power Deliv.} \textbf{11}(3) (1996) 1238.

\bibitem{theory.fbd}	
M. Depenbrock, \emph{IEEE Trans. Power Syst.} \textbf{8}(2) (1993) 381.

\bibitem{power.epe}
A. Costabeber, P. Tenti, T. Caldognetto, and E. Verri Liberado, Selective compensation of reactive, unbalance, and distortion power in smart grids by synergistic control of distributed switching power interfaces, Proc. 15th European Conf. on Power Electronics and Applications (EPE), 2013.	

\bibitem{power2.epe}
R.S. Herrera, P. Salmeron, J.R. Vazquez, S.P. Litran, and A. Perez, Generalized instantaneous reactive power theory in poly-phase power systems, Proc. 13th European Conf. on Power Electronics and Applications (EPE '09), 2009.
	
\bibitem{theory.kim}
H. Kim, F. Blaabjerg, B. Bak-Jensen, and J. Choi, \emph{IEEE Trans. Power Electron.}  \textbf{17}(5) (2002) 701.

\bibitem{theory.Czarnecky}
L.S. Czarnecki, \emph{IEEE Trans. Power Deliv.} \textbf{21}(1) (2006 362.

\bibitem{theory.Herrara}
R.S. Herrera and P. Salmeron, \emph{IEEE Trans. Power Deliv.} \textbf{22}(1) (2007) 595.

\bibitem{theory.tenti}
P. Tenti, H.K.M. Paredes, and P. Mattavelli, \emph{IEEE Trans. Power Electron.} \textbf{26}(3) (2011) 664.

\bibitem{theory.suh}
Y. Suh and T.A. Lipo, \emph{IEEE Trans. Power Deliv.} \textbf{21}(3) (2006) 1530.


\bibitem{compensation.epe}
P. Tenti, D. Trombetti, E. Tedeschi, and P. Mattavelli, Compensation of load unbalance, reactive power and harmonic distortion by cooperative operation of distributed compensators, Proc. 13th European Conf. on Power Electronics and Applications (EPE '09), 2009.

\bibitem{power.song}
H.S. Song and K. Nam, \emph{IEEE Trans. Ind. Electron.} \textbf{46}(5) (1999) 953.

\bibitem{power.suh}
Y. Suh and T.A. Lipo, \emph{IEEE Trans. Ind. Appl.} \textbf{42}(3) (2006) 825.

\bibitem{power.siemaszko}
D. Siemaszko and A. Rufer, Power compensation approach and double frame control for grid connected converters, Proc. 10th IEEE International Conf. on Power Electronics and Drive Systems (PEDS), 2013.

\end{thebibliography}
\end{document}